\documentclass[twocolumn,showpacs,preprintnumbers,amsmath,amssymb,prb]{revtex4}
\usepackage{graphicx}
\usepackage{color}
\usepackage[english]{babel}
\usepackage[T1]{fontenc}
\usepackage[latin1]{inputenc}
\usepackage{amsmath}
\usepackage{amssymb}
\usepackage{subfigure}

\def \ie{{\it\frenchspacing i.e.\ }}

\def\d{\mbox{d}}

\def     \a{\`{a}~}

\def     \pl{\partial}

\begin{document}

\title{Direct Numerical Simulation of Turbulent Heat Transfer 
Modulation in Micro-Dispersed Channel Flow
\footnote{
It is our great pleasure to take part in this Festschrift Issue
dedicated to Professor Franz Ziegler on the occasion of his 70th
birthday. This study was crafted with friendship and respect to honour
his activity and his scientific achievements. We wish Franz many more
productive, enjoyable and happy years and a solid and long collaboration
as Editors of Acta Mechanica.
}}

\author{Francesco Zonta, Cristian Marchioli and Alfredo Soldati\thanks{Author to whom
correspondence should be addressed E-mail: soldati@uniud.it~~Phone: +39~(0432)~558020.
Also affiliated with the Department of Fluid Mechanics, CISM, 
33100, UDINE, Italy.}}
\address{
Centro Interdipartimentale di Fluidodinamica e Idraulica and
Dipartimento di Energetica e Macchine, Universit\a degli Studi di Udine,
33100, Udine, Italy}


\begin{abstract}
{\bf Abstract} The object of this paper is to study the influence of dispersed micrometer
size particles on turbulent heat transfer mechanisms in wall-bounded flows.
The strategic target of the current research is to set up a methodology to 
size and design new-concept heat transfer fluids with properties 
 given by those of the base fluid modulated by the presence 
of dynamically-interacting, suitably-chosen, discrete micro- and nano- particles.

We run Direct Numerical Simulation (DNS) for hydrodynamically fully-developed,
thermally-developing turbulent channel flow at shear Reynolds number $Re_{\tau}=150$ 
and Prandtl number $Pr=3$, and we tracked two large swarms of particles,
characterized by different inertia and thermal inertia.
Preliminary results on velocity and temperature statistics for both phases
show that, with respect to single-phase flow, heat transfer fluxes
at the walls increase by roughly $2 \%$ when the flow is laden with
the smaller particles, which exhibit a rather persistent
stability against non-homogeneous distribution and near-wall
concentration. An opposite trend (slight heat transfer flux decrease)
is observed when the larger particles are dispersed into the flow.
These results are consistent with previous experimental findings and are 
discussed in the frame of the current research activities in the field. 
Future developments are also outlined.  
%

\end{abstract}

\maketitle

\section{Introduction}
The problem of developing efficient heat transfer techniques
for technological applications
has become more and more important over the last decades due
to the increasing demand of cooling in high heat flux equipments
and to the unprecedented pace of component miniaturization
[\ref{ref1},\ref{ref2},\ref{ref3}].
Consider, for instance, teraflop computers and other electronic equipments
like optical fibers, high energy density lasers and high power x-rays.
These devices are required to operate
with high precision and, at the same time, with minimum size.
Such requirements impose
a challenge in terms of both device design and thermal management,
not only in the case of micro-scale applications but
also for large-scale applications such as 
transport vehicle engines, fuel cells and controlled bio-reactors [\ref{ref3}]. 
Common air-based cooling systems have proven inadequate in high
heat flux applications and more efficient techniques for heat
transferability are thus required.
In particular, a quest for a fluid with high heat transfer capacity,
characterized by the possibility of tunable thermal properties and
also associated to
low management/safety problems has started.
One possibility is to use nanofluids, namely dilute liquid suspensions
of nanoparticulate solids including particles, nanofibers and nanotubes,
which are supposed to change the heat transfer capabilities of the base fluid
up to the target, desired amount.
Nanofluids were
first brought into attention approximately a decade ago, when their enhanced
thermal behavior was observed with respect to conventional single-phase
fluids such as water, engine oil, and ethylene glycol [\ref{ref3}].
Specifically, due to the high thermal conductivity
of metals in solid form, fluids containing suspended metal particles
display significantly enhanced thermal conductivity and specific heat capacity
[\ref{Xuan},\ref{Maiga_palm}] compared to the
conventional heat transfer liquids.
For example,
the thermal conductivity of gold at room temperature is more than 500 times larger than that
of water and more than 2000 times larger than that of engine oil.

The idea of increasing the effective thermal conductivity of fluids with
suspensions of solid particles is not so recent since
the first theoretical formulation of nanofluids as a new concept of heat
transfer fluids was put forth by Maxwell [\ref{maxwell}] more than one century ago,
then followed by several other experimental and theoretical studies, such as those of
Hamilton and Crosser [\ref{Hamilton}] and Wasp et al. [\ref{Wasp}].
As of today, however, a clearcut understanding of the modifications of the
heat transfer mechanisms occurring in nanofluids is still to be produced.
This is indicated by the failure of classical models of
suspensions and slurries in predicting nanofluids 
behavior [\ref{ref3},\ref{Wongwises}] accompanied by the lack of
alternative modelling strategies.
Such lack may be possibly due to the number of investigations on the physical
mechanisms which govern the heat transfer processes, still relatively
low if compared to the corrispondingly large number of
theoretical and experimental studies devoted to
the problem of enhancing heat transferability with nanofluids
(see [\ref{ref3},\ref{Wongwises}] for a review).
A possible way to improve the understanding of the physical
transfer mechanisms is to use accurate and reliable numerical
tools such as Direct Numerical Simulation (DNS)
and Lagrangian Particle Tracking (LPT), which may complement
complex and costly experiments.
DNS-based Eulerian-Lagrangian studies have been widely used for
investigating mass, momentum and heat transfer mechanisms
in turbulent boundary layers.
In particular, previous DNS studies on turbulent particle dispersion in
wall-bounded flows (see [\ref{ms02}-\ref{ef94}]
among others) have proven their capability of predicting particle-turbulence
interactions: these studies made it possible to perform  phenomenological and 
quantitative analyses on the
dispersion processes [\ref{s05},\ref{bee}], highlighting clustering and segregation
phenomena. In the process of transferring this research methodology
to nanofluids, one question requiring clarification
for clearcut expectations is to
what extent the same approach is proper to solve
the different new physics challenges imposed by the different phenomena
and range of parameters.

In this work, we propose to start approaching nanofluids from
a simplified simulation setting where microparticles,
rather than nanoparticles, are dispersed into the flow.
This simplification allows to single out complications
arising when the particles are very small and
compare with the molecular scales of the fluid: inter-particle
forces, wall effects, van der Waals forces, Brownian
diffusion, etc.
DNS has been used already to investigate on turbulent heat transfer in wall-bounded flows
[\ref{Chagras}-\ref{lyons}]
and available studies provide systematic analyses of the Reynolds and Prandtl
number effects on the heat transfer process. However, all these studies
consider either single-phase
turbulent flows [\ref{Lakehal_ful},\ref{tiselj}-\ref{lyons}]
or turbulent flows laden with coarse particles [\ref{Chagras},\ref{hetsroni}].
A comprehensive analysis
accounting for mass, momentum and heat transfer mechanisms all together
and tailored for the specific case of micro- or nano-dispersed fluids is
currently unavailable as far as our knowledge goes.
This is due to the non-trivial modeling issues,
which of course reflect upon the complex interactions between the two phases.
To elaborate, studying heat transfer modifications requires modeling
particles as active heat transfer agents which interact both with
the temperature field and the velocity field.
Necessary energy and momentum coupling terms must be
incorporated in the governing equations of both phases.
This paper represents an effort toward a systematic phenomenological study
of turbulent heat transfer mechanisms in micro- and/or nano-dispersed fluids.
Since the modeling of these fluids represents
a largely unexplored field of research, this study involves substantial challenges
due to the rich complexity of the involved physics. The main focus of the
present paper is to examine the modifications
produced by solid inertial particles on the temperature fields of both fluid and particles.
First, the numerical methodology that we use to investigate on the problem will be
described; then preliminary statistical results will be shown and discussed
in the limit of hydrodynamically fully-developed, thermally developing turbulent channel flow
laden with particles large enough to neglect Brownian diffusion
(which becomes important only for particle diameters smaller than $1~\mu m$)
but small enough to
ensure stability against the inertia-dominated non-homogeneous distribution
[\ref{ef94}] and consequent near-wall accumulation [\ref{ms02},\ref{s05}].

\section{Methodology}
\subsection{Flow field equations} 
\begin{figure}[t]
\centerline{\hbox{\includegraphics[width=8.cm,angle=270]{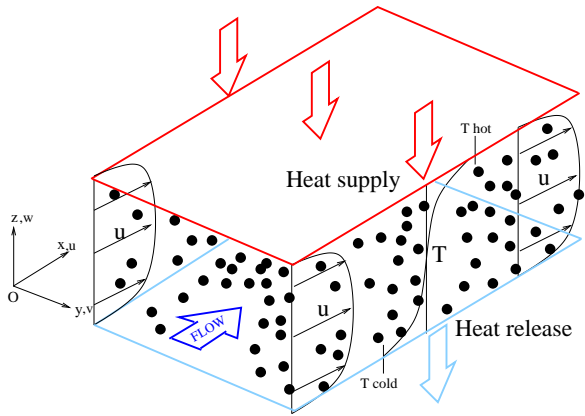}}}
\vspace{-1.2cm}
\caption{Sketch of the computational domain.}
\label{sketch}
\end{figure}

With reference to the schematics of Fig.
\ref{sketch}, particles are introduced in a turbulent channel
flow with heat transfer. Assuming
that the fluid is incompressible and Newtonian,
the governing balance equations for the fluid (in
dimensionless form) read as:
\begin{equation}
\label{mass_adim}
\frac{\partial{u_i}}{\partial{x_i}} = 0~,
\end{equation}
\vspace{-0.3cm}
\begin{equation}
\label{ns_adim}
\frac{\partial{u_i}}{\partial{t}} = -u_{j}\frac{\partial{u_i}}{\partial{x_j}} +
\frac{1}{Re^*}\frac{\partial^{2}{u_i}}{\partial{x_j}^2} - \frac{\partial{p}}{\partial{x_i}}
+ \delta_{1,i}~ + f_{2w},
\end{equation}
\vspace{-0.3cm}
\begin{equation}
\label{en_adim}
\frac{\pl T}{\pl t}+ u_j\frac{\pl T}{\pl x_j}= 
\frac{1}{Re^* Pr}\frac{\pl^2 T}{\pl x^2_{j}} + q_{2w},
\end{equation}
where $u_i$ is the $i^{th}$ component of the velocity
vector, $p$ is the fluctuating kinematic pressure, $\delta_{1,i}$
is the mean pressure gradient that drives the flow, 
$T$ is the temperature, $Re^*$ is the shear (or friction)
Reynolds number and $Pr$ is the Prandtl number.
The shear Reynolds number is defined as $Re^* = u^*h/\nu$,
based on the shear velocity, $u^*$, on the half channel
height, $h$, and on the fluid kinematic
viscosity, $\nu$.
The shear velocity is defined as $u^* = (\tau_{w}/\rho)^{1/2}$,
where $\tau_{w}$ is the mean shear stress at the wall
and $\rho$ is the fluid density.
The Prandtl number is defined as $Pr=\mu c_p / k$ where
$\mu$, $c_p$ and $k$ are the dynamic viscosity,
the specific heat and the thermal conductivity of the fluid, respectively.
All variables considered in this study are reported in dimensionless
form, represented by the superscript +, which has been dropped from Eqns.
(\ref{mass_adim}) to (\ref{en_adim}) for ease of reading, and expressed
in wall units. Wall units are obtained by taking $u^*$, $\nu$ and
the shear (or friction) temperature $T^*$ as the reference quantities
employed for normalization. The shear temperature is defined
as $T^*= q_w/\rho c_p u^*$ where $q_w=k \cdot \nabla T_w$
is the mean heat flux at the wall and $\nabla T_w$ is the wall-normal
component of the temperature gradient at the wall.
In Eqns. (\ref{ns_adim}) and (\ref{en_adim}) the momentum-coupling term
$f_{2w}$ and the energy-coupling term $q_{2w}$ are defined in terms of
momentum and energy flux per unit mass, namely $f_{2w}=F_{2w}/m_p$ and
$q_{2w}=Q_{2w}/(c_p \cdot m_p)$ where $m_p$ is the particle mass.
These terms are introduced to
model, as point sources, the influence of the particles on 
the fluid velocity and temperature fields (two-way coupling approach).
More details on these terms are given in Sec. \ref{par2way}.
The reference geometry consists of two infinite
flat parallel walls; the origin of the coordinate system is located
at the center of the channel and the $x-$, $y-$ and $z-$ axes point
in the streamwise, spanwise and wall-normal directions respectively
(see Fig.~\ref{sketch}).
The calculations are performed
on a computational domain of size
$4 \pi h \times 2 \pi h \times 2 h$
in $x$, $y$ and $z$, respectively.
Periodic boundary conditions are imposed on both velocity and temperature
in the homogeneous $x$ and $y$ directions; at the wall, no-slip condition is enforced
for the momentum equation whereas constant temperature condition is 
adopted for the energy equation. Specifically, the temperature on the
boundaries is held constant at the uniform values $<T_w^h>=80^o~C$ for
the {\em hot} wall (source of heat) and $<T_w^c>=20^o~C$ for the {\em cold}
wall (sink of heat), where $<>$ denotes averaged values.
The condition on temperature has been chosen because our aim is to 
simulate the problem of thermally-developing forced convection in a
channel where supply of heat from a source
and release of heat to a sink are considered as
constant temperature processes.

\subsection{Particle equations} 
\label{parteq}

Large samples of heavy particles with diameter $d_p=4~\mu m$ and $8~\mu m$
and with density $\rho_p=19.3 \cdot 10^3~kg~m^{-3}$
(gold in water) are injected into the flow at concentration high enough to
have significant two-way coupling effects in both the momentum and energy
equations but negligible particle-particle interactions.
The particle dynamics is described by a set of ordinary differential equations
for position, velocity and temperature.
For particles heavier than the fluid ($\rho_{p}/\rho \sim 20$,
as in the present case), Armenio and Fiorotto [\ref{armenio}]
showed that the most significant force is Stokes drag.
Other forces acting on the particles, such as hydrostatic force, Magnus effect,
Basset history force and added mass force are not taken into account since they
are assumed to be negligible (orders of magnitude smaller) because
of the specific set of physical parameters of our simulations [\ref{armenio},\ref{re}].
We also neglected the Brownian force, which is proportional to  $1/d_p^5$ and
becomes important for particle diameters less than 1 $\mu m$ [\ref{Soltani}].
With the above assumptions, a simplified version of the Basset-Boussinesq-Oseen equation
for particle momentum balance [\ref{crowebook}] is obtained. In vector form:
\begin{eqnarray}
\label{partpos}
\frac{d{\bf x_p}}{dt} = {\bf u_p}~,
\end{eqnarray}
\begin{equation}
\label{mom_part}
\frac{d{\bf{u}_p}}{dt}=- \frac{3}{4} \frac{C_D}{d_p} \left( \frac{\rho}{\rho_p}\right) \mid \bf{u}_p-u \mid \left( \bf{u}_p-u \right)~,
\end{equation}
where ${\bf x_p}$ is the particle position,
$\bf{u}_p $ is the particle velocity, $ \bf{u} $ is the fluid velocity
at particle position and $C_D$ is the drag coefficient.
The formulation of the drag coefficient $C_D$ follows the non-linear approximation reported in 
Schiller and Naumann [\ref{Schiller}]:
\begin{equation}
\label{c_d}
C_D= \frac{24}{Re_p} \left( 1 + 0.15Re_p^{0.687} \right)~,
\end{equation}
where $Re_p=d_p \mid \bf{u}_p - u \mid / \nu$ is the particle Reynolds number
based on the relative particle-to-fluid velocity.
The correction for $C_D$ is necessary when $Re_p$ does not remain small ($Re_p>1$).

In this study, the particles exchange both momentum and heat with the carrier fluid.
The equation describing the temperature evolution of the dispersed phase can be derived
from the energy balance on a particle, written under the assumption of convective
heat transfer occurring through the particle surface (the contribution of radiation
is thus neglected). The inter-phase heat transfer rate of a spherical particle in motion
relative to the surrounding fluid can be written as:
\begin{equation}
\label{flux_part}
\dot{Q_c}= Nu \cdot \pi d_p \cdot k_f \cdot \left( T_s-T \right)~,
\end{equation}
where $Nu$ is the Nusselt number, $k_f$ is the thermal conductivity of the
fluid, $T$ is the temperature of the fluid and $T_s$ is the temperature at 
the particle surface. The Nusselt number is given by the well-known Ranz-Marshall
correlation [\ref{rm}]:
\begin{equation}
\label{Ranz_Mar}
Nu= 2+ 0.6 \cdot {Re_p}^{\frac{1}{2}} \cdot {Pr}^{\frac{1}{3}}~,
\end{equation}
and accounts for changes of the heat transfer rate due to the relative motion between
the particle and the surrounding fluid.
Rearranging Eqn. (\ref{flux_part}), one can write:
\begin{equation}
\label{flux_part_new}
\frac{dT_p}{dt} = \frac{Nu}{2} \cdot \frac {\left( T_f-T_p \right)}{\tau_T}~,
\end{equation}
where $T_f$ is the temperature of the fluid at particle position, $T_p$
is the temperature of the particle
and $\tau_T=(c_p \rho_p {d_p}^2) / (12 k_f )$ is the particle thermal response time.
In this study, we considered uniform temperature inside the particle, namely
$T_p=T_s$: this assumption is
justified for particles with Biot number smaller than 0.1 [\ref{crowebook}].
The particle Biot number is defines as: 
\begin{equation}
\label{Biot}
Bi=\frac{h_p d_p} {2 \lambda}~,
\end{equation} 
where $h_p$ and $k_p$ are the convective heat transfer coefficient and the thermal conductivity
of the particle. In the present simulations, the particle Biot number is
${\cal{O}} (10^{-6})$.

\subsection{Modeling of two-way coupling}
\label{par2way}

The source terms $f_{2w}$ and $q_{2w}$ in Eqns. (\ref{ns_adim}) and (\ref{en_adim})
arise because of the momentum transfer due to the drag force on the particle
and because of the convective heat transfer to/from the particle, respectively.
Calculation of these coupling terms is done applying the
action-reaction principle to a generic volume of fluid $\Omega$ containing a particle.
Focusing on the $f_{2w}$ term (the extension for the term
$q_{2w}$ is straightforward), we have:
\begin{equation}
\label{act_react}
\int_{\Omega} {\bf f}_{2w}({\bf x}) d\Omega=-{\bf f}_{flu}~,
\end{equation} 
where ${\bf f}_{flu}$ is the force exerted by the fluid on the particle. 
The term ${\bf f}_{2w}$, which represents the feedback of the dispersed
phase on the fluid, can be obtained by adding the contributions of each particle:
\begin{equation}
\label{sum_f}
{\bf f}_{2w}=\sum_{p=1}^{n_p} \left( {\bf f}_{2w}^{p} \right)~,
\end{equation} 
where $n_p$ is the total number of particles.
The evaluation of each contribution ${\bf f}_{2w}^{p}$ is obtained
using the point source approximation [\ref{Boivin},\ref{Sundaram}]:
\begin{equation}
\label{point_source}
{\bf f}_{2w}^{p}= - {\bf f}_{flu} \delta ({\bf x}-{\bf x}_p)~,
\end{equation}
where $\delta ({\bf x}-{\bf x}_p)$ is the Dirac delta function.

\subsection{DNS methodology}
\label{DNSsimul}

In this study a DNS of fully-developed channel flow with heat transfer is performed.
The governing equations for the fluid, Eqns. (1)-(3), are discretized
using a pseudo-spectral method
based on transforming the field variables into wavenumber space, using Fourier
representations for the periodic (homogeneous) directions and a Chebyshev representation for the
wall-normal (non-homogeneous) direction.
As commonly done in pseudospectral methods, the convective non-linear terms are first computed
in the physical space and then transformed in the wavenumber space using a de-aliasing procedure
based on the 2/3-rule; derivatives are evaluated directly in the wavenumber space to maintain spectral accuracy.
Time advancement of the equations is performed using an explicit two-stage Euler/Adams-Bashforth scheme
for convective terms and an implicit Crank-Nicolson method for the viscous terms.
The time step used is $dt^+=0.045$ in wall units.
More details on the numerical scheme can be found in [\ref{Lam1}].

DNS calculations were performed at $Re^*=150$,
corresponding to a bulk (average) Reynolds number $Re_{b} = u_b h / \nu = 1900$,
where $u_b$ is the bulk velocity.
At this Reynolds number, two different values of the Prandtl number
were considered (see Table \ref{summarysimul}).
\begin{table}[b]
\centering
\begin{tabular}{c c c c c c c c c c}
\hline
\hline
RUN & $Re^*$ & $u^*$ & $Pr$ & $d_p$     & $d_p^+$ & $\tau_p$ & $St$ & $\tau_T$ & $St_{T}$ \\
    &        & [m/s] &      & $[\mu m]$ &         & $[\mu s]$    & $(=\tau_p^+)$ & $[\mu s]$    & $(=\tau_T^+)$ \\ \hline
$R1$ & 150 & 0.11775 & 0.71 & --- & --- & --- & --- & --- & --- \\ \hline
$R2$ & 150 & 0.3 & 3 & --- & --- & --- & --- & --- & --- \\ \hline
$R3$ & 150 & 0.3 & 3 & 4 & 1.2 & $17.16 $ & $ 1.56 $ & $5.49$ & $ 0.5 $ \\ \hline
$R4$ & 150 & 0.3 & 3 & 8 & 2.4 & $68.62 $ & $ 6.24 $ & $2.196$ & $ 2.0 $ \\
\hline
\hline
\end{tabular}
\vspace{-0.1cm}
\caption{Summary of the simulation parameters.}
\label{summarysimul}
\end{table}
First, a simulation at $Pr=0.71$ (run $R1$ in Table \ref{summarysimul}) was performed
to validate the flow solver
against numerical data available in the literature [\ref{Kasagi}] for the situation
of turbulent flow of air ($\rho=1.3~kg~m^{-3}$, $\nu = 15.7 \cdot 10^{-6}~m^2/s$)
at ambient temperature in a channel of half height $h=0.02~m$. In this simulation, we
have $u^*=0.11775~m~s^{-1}$ and $u_b = 1.49~m~s^{-1}$.
Second, simulations at $Pr=3$ (corresponding to runs $R2$ to $R4$ in Table \ref{summarysimul})
has been performed to study heat transfer modifications
in a particle-laden turbulent flow of water ($\rho=10^3~kg~m^{-3}$, $\nu = 10^{-6}~m^2/s$)
at the temperature $<T>=50^{o}$ C in a channel of half height $h=500~\mu m$.
The temperature $<T>$ corresponds
to the mean temperature between $<T_w^h>$ and $T<_w^c>$.
In this case, $u^*=0.3~m~s^{-1}$ and $u_b = 3.8~m~s^{-1}$.

The computational domain has dimensions $1885\times942\times300$ in wall units
and was discretized using an Eulerian grid made of $128\times128\times129$
nodes (corresponding to $128\times128$ Fourier modes and to 129 Chebyshev
coefficients in the wavenumber space).
The grid spacing is uniform in the homogeneous directions
and corresponds to spatial resolutions equal to $\Delta x^+ =14.72$
and $\Delta y^+ =7.35 $; the nodes along the wall-normal direction are
clustered near the wall corresponding to spatial resolutions
from $\Delta z^+ =0.0452$ at the wall to $\Delta z^+ =3.68 $
at the centerline.
The wall-normal grid spacing is always smaller than the smallest
local flow scale and, thus, it fulfills the requirements imposed
by the point-particle approach (see Sec. IIE).

\subsection{Lagrangian particle tracking}

The complete set of equations which describes the time evolution of particle position, 
velocity and temperature in the turbulent flow field is given by Eqns. (\ref{partpos}),
(\ref{mom_part}) and (\ref{flux_part_new}).
To solve for these equations, we have coupled the DNS flow solver to a Lagrangian
tracking routine.
The routine uses $6^{th}$-order Lagrangian polynomials to interpolate the fluid velocity components
and the fluid temperature at particle position.
The performance of the interpolation scheme is comparable to that of spectral
direct summation and to that of an hybrid scheme which exploits $6^{th}$-order Lagrangian polynomials in
the streamwise and spanwise directions and Chebyshev summation in the wall-normal direction.
A $4^{th}$-order Runge-Kutta scheme is used for time
advancement of the particle equations.
The timestep size is equal to that used for
the fluid ($\delta t^+ = 0.045$). The total tracking time
was $t^+ = 2000$ for the $4~\mu m$ particles and
$t^+ = 1500$ for the $8~\mu m$ particles. We remark here that
these simulation times are not sufficient to achieve a statistically
steady state for the particle concentration, yet they are long enough
to obtain converged velocity and temperature statistics and to
highlight qualitatively the effect of the particles on the heat
transfer rate.
Particles are treated as pointwise, rigid spheres (point-particle approach)
and are injected into the flow at average mass fraction, $\Phi_m$,
high enough to have a two-way
coupling between the particles and the fluid ($\Phi_m \sim 10^{-2}$)
[\ref{Uijttewaal},\ref{Kiger}].
Possible effects due to inter-particle collisions
are neglected.
At the beginning of the simulation, particles
are distributed randomly over
the computational domain and their initial
velocity and temperature are set equal to those of the fluid
at the particle initial position.
Periodic boundary conditions are imposed on particles
moving outside the computational domain in the
homogeneous directions, whereas
perfectly-elastic collisions at the smooth walls are assumed
when the particle center is less than one
particle radius away from the wall.
No specific boundary condition is needed for the particle temperature equation
(Eqn. \ref{flux_part_new}) since the integration of this equation follows the
integration of the particle momentum equation (Eqn. \ref{mom_part})
and it requires only the knowledge of the initial condition.
In the simulations presented here,
large samples of $800,000$ particles have been
considered for each value of $Re^*$ and $Pr$.
We remark here that tracking of ${\cal{O}}~(10^6)$ particles
two-way coupled with the fluid
requires a huge computational effort
in terms of both computational cost of the simulation
and disk storage availability, considering also the
rather long tracking times achieved in the simulations.
Each particle sample is characterized by
different values of the particle response
times. Table ~\ref{summarysimul} summarizes the complete set of
parameters relevant to the simulations of particle dispersion,
including the particle Stokes numbers, $St$ and $St_T$.
The particle Stokes number corresponds to the non-dimensional particle response
time and is obtained using the viscous timescale $\tau_f=\nu/u_{\tau}^2$
as reference. In the present study, we have $St=\tau_p^+=\tau_p/\tau_f$
and $St_T=\tau_T^+=\tau_T/\tau_f$.

\section{Results and discussion}
\subsection{Unladen turbulent channel flow with heat transfer}
\label{unladen}

In this paragraph we examine the statistics relative to velocity and thermal variables
for the base case of unladen fluid. We examine the results relative to the 
simulations $R1$ and $R2$, performed at 
the same Reynolds number ($Re^*= 150$) and at Prandtl numbers equal to $0.71$ and $=3$,
respectively. 
Velocity and temperature statistics will be examined
and compared against literature reference cases [\ref{na_han},\ref{Kasagi}].

\subsubsection{Velocity field}
\begin{figure}[b]
\centerline{\hbox{\includegraphics[width=6.5cm,angle=270]{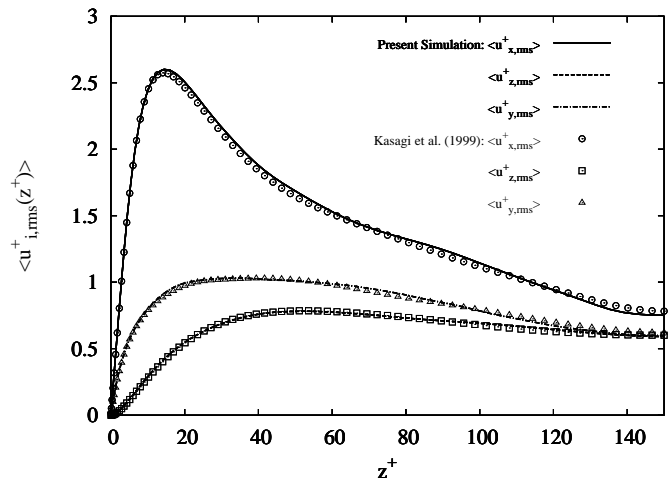}}}
\vspace{-0.2cm}
\caption{Rms of fluid velocity components, $<u^+_{i,rms}(z^+)>$, for
single-phase (unladen) turbulent channel flow. }
\label{velrms1way}
\end{figure}

The mean velocity profile will not be shown since it collapses onto the logarithmic law of the wall
perfectly [\ref{s05}] and matches previous results obtained in Refs. [\ref{na_han},\ref{Kasagi}].
In Fig. (\ref{velrms1way}) the root mean square (rms) of the fluctuations of the velocity components,
$<u^+_{i,rms}(z^+)>$, are plotted as a function of the
wall normal coordinate in wall units, $z^+$,
and compared against the results of Kasagi and Iida [\ref{Kasagi}].
The agreement is generally good showing small differences which may be due to marginal statistical
sampling of the time series. 
In the two simulations relative to the base fluid only, 
the velocity field depends only on the pressure gradient (namely on the shear Reynolds number, $Re^*$)
and it is not influenced by the value of the Prandtl number -- forced convection. 

\subsubsection{Temperature field}
\begin{figure}[t]
\centerline{\hbox{\includegraphics[width=6.5cm,angle=270]{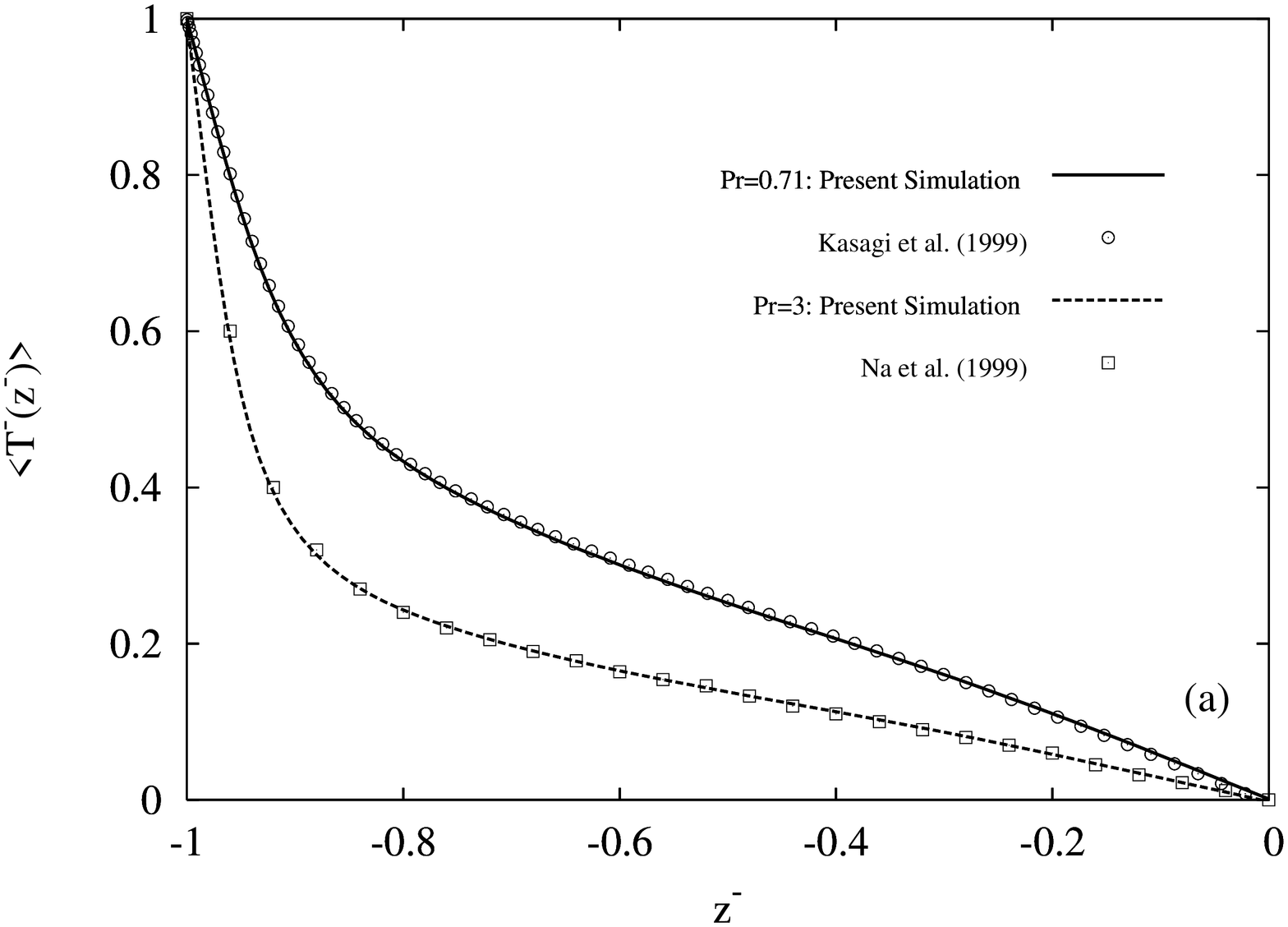}}}
\centerline{\hbox{\includegraphics[width=6.5cm,angle=270]{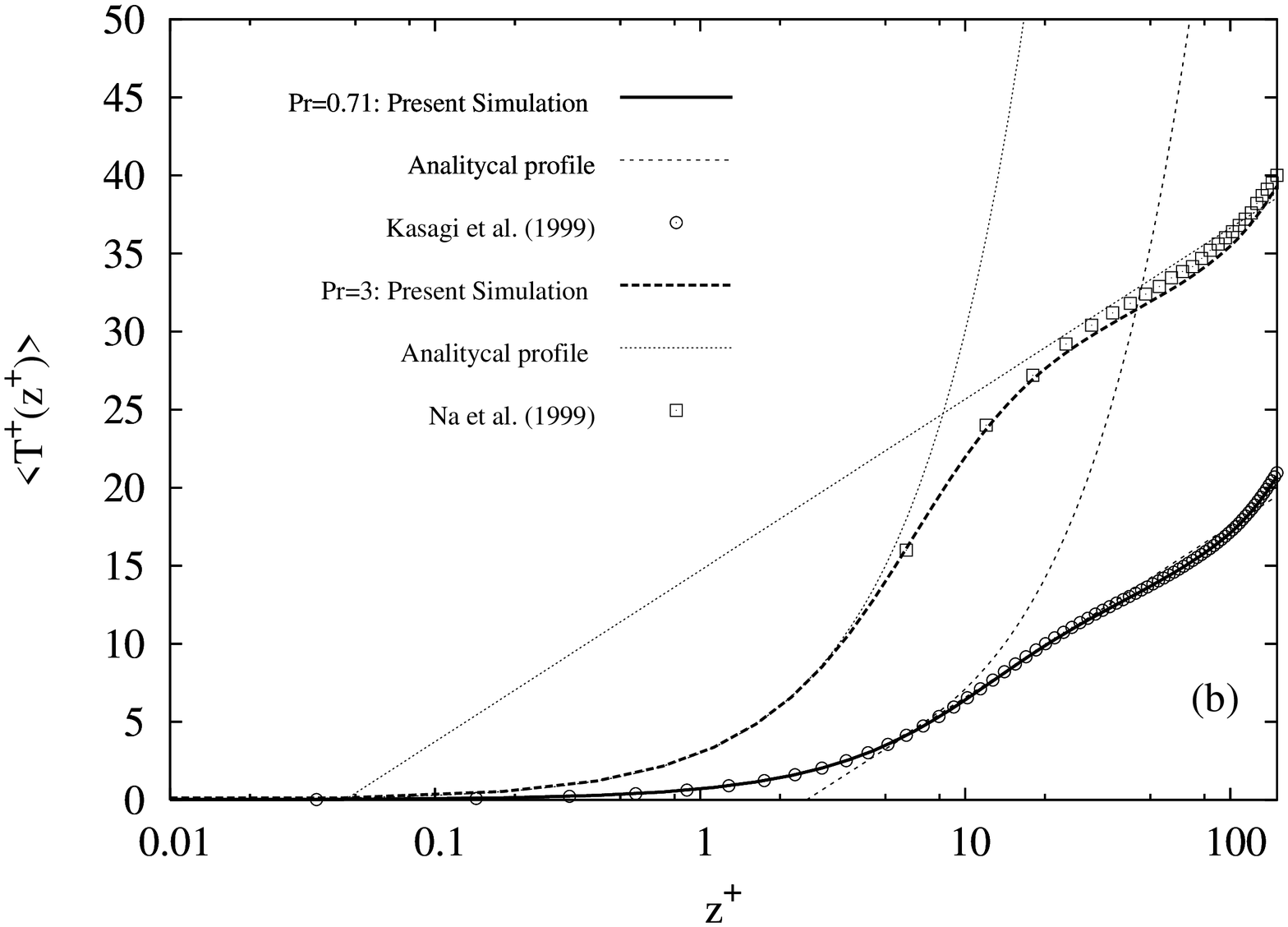}}}
\vspace{-0.2cm}
\caption{Mean fluid temperature at $Pr=0.71$ and $Pr=3$. Panels: (a) $<T^-(z^-)>$ (linear-linear scale),
(b) $<T^+(z^+)>$ (log-linear scale).}
\label{tmean071}
\end{figure}

The behavior of the fluid temperature averaged over the homogeneous directions ($x$ and $y$) is shown in
Fig. (\ref{tmean071}) for the two values of the Prandtl number, $Pr=0.71$ and $Pr=3$ respectively.
In Fig. (\ref{tmean071}a), the temperature is made dimensionless in outer
units, indicated by the superscript --, as follows:
\begin{equation}
<T^->=\frac{ <T> - T_m}{\Delta T_m}~,
\end{equation}
where $T_m= (T_H+T_C)/2$ is the average centerline temperature
and $\Delta T_m= (T_H-T_C)/2$ is the temperature difference between the walls. 
The average temperature is shown as a function of the wall normal coordinate 
made dimensionless by the half channel height, $h$.
The results of the present work are compared and assessed against those computed by Kasagi and Iida [\ref{Kasagi}] 
 for $Pr=0.71$ 
and against those computed by Na et al. [\ref{na_han}] for
$Pr=3$. Results match perfectly showing that the simulation parameters
-- \ie resolution, average time, etc. --
are appropriate for the problem under investigation.

The same averaged temperature profiles are plotted in Fig. (\ref{tmean071}b)
using a semi-logarithmic scale just to
show possible minor differences among the current profiles and the benchmark profiles 
[\ref{na_han},\ref{Kasagi}].
Temperature profiles are shown as a function of the wall normal distance, both being
expressed in wall units.
In this case, the temperature is defined as the difference between the local average
temperature and the wall temperature (both relative
to the average centerline temperature) normalized by the friction temperature,
$T^*=q_w / (\rho_f c_p u^*)$: $<T^+>=(<T>-{T_W})/T^*$.

The computed data are also compared with standard analytical correlations used to estimate
the temperature wall dependence. These correlations have the following form:
\begin{equation}
<T^+> = \frac{<\overline{T}>-T_W}{T^*}=
\begin{cases}
Pr \cdot z^{+} & \text {if $z^{+}\leq 11.6$ }, \\
\frac{1}{k_{\theta}} \ln z^+ + B_{\theta} & \text {if $z^{+}>11.6$} 
\end{cases} 
\label{a} 
\end{equation}
where $k_{\theta}=0.21$ and $B_{\theta}=-4.4$ and $14.7$ for $Pr=0.71$ and for $Pr=3$, respectively.

As broadly known, the profiles show the existence of a (near-wall) diffusive sublayer,
the thickness of which varies with the Prandtl number and is approximately equal 
to $\Delta \theta^+=6.5$ for $Pr=0.71$ and $\Delta \theta^+=4.5$ for $Pr=3$.
In Figs. (\ref{tmean071}a) and (\ref{tmean071}b) we can further 
observe that the temperature gradient at the wall is strongly 
dependent on the value of the Prandtl number. This behavior is shown by Eqn.
(\ref{a}), from which it is clear that the temperature in the viscous sublayer 
depends linearly on the Prandtl number.

In problems which involve turbulent heat transfer, the Prandtl number is also 
important to establish the smallest spatial scale for the 
temperature field, $\eta_{\theta}$, which 
can be expressed [\ref{Monin},\ref{Batchelor}] as a function of
the Kolmogorov scale, $\eta_{k}$, as follows:
\begin{equation}
\label{pr_min_1}
\eta_{\theta} \sim \eta_{k} { \left ( \frac{1}{Pr} \right ) }^{3/4}~,
\end{equation}
for $Pr<1$, and:
\begin{equation}
\label{pr_magg_1}
\eta_{\theta} \sim \eta_{k} { \left ( \frac{1}{Pr} \right ) }^{1/2}~,
\end{equation}
for $Pr>1$. The previous relations confirm that 
for a given value of $\eta_{k}$ (we remind here that the average value
of $\eta_{k}$ depends only on the Reynolds number, which 
is equal to 150 in the present simulations), the smallest temperature scale decreases for increasing 
Prandtl numbers. 

\begin{figure}[h]
\centerline{\hbox{\includegraphics[width=6.5cm,angle=270]{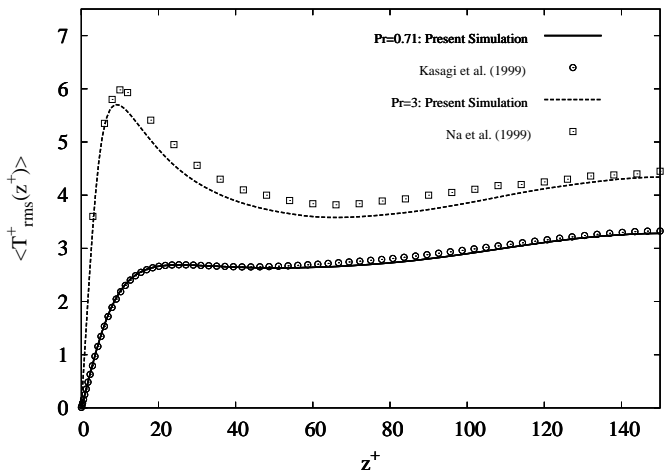}}}
\vspace{-0.3cm}
\caption{Rms of fluid temperature, $<T^+_{rms}(z^+)>$, at $Pr=0.71$ and at $Pr=3$.}
\label{trms071}
\end{figure}

The behavior of rms of the temperature field fluctuations,
$<T^+_{rms}(z^+)>$, is shown in Fig. (\ref{trms071}). Results
obtained by Kasagi and Iida [\ref{Kasagi}] for $Pr=0.71$ and by Na et al. [\ref{na_han}] for $Pr=3$ are also shown
for benchmark and comparison. Results are presented in wall variables. 
Focusing on the $Pr=0.71$ case, we observe that the temperature intensity reaches a maximum at the channel centerline, 
not mimicking the
behavior of the fluctuations of the velocity field which all reach their peak in the wall 
proximity (see Fig. (\ref{velrms1way})). This was
discussed by Lyons et al. [\ref{lyons}], who attributed this different behavior
to the temperature boundary conditions 
which force a non-zero temperature gradient at the center of the channel.
In the $Pr=3$ case, the temperature intensity reaches a maximum in the near-wall region. 
This observation is related to the Prandtl number effect on $<T^+_{rms}(z)>$,
indicating that the range of wavenumbers in the thermal fluctuating field increases with $Pr$,
for which the spectral functions of the velocity fields are negligible.
As can be also observed, the increase in the Prandtl number corresponds to a
shift of the peak value of the temperature fluctuations toward the wall. 
Both in the current results and in those by Kasagi and Iida [\ref{Kasagi}] for $Pr=0.71$, the 
relative peak of the temperature fluctuations is located around $z^+=20$.
The location of the peak for the higher Prandtl number ($Pr=3$) has moved to $z^+=9$, roughly.
Small discrepancies due perhaps to  the marginally statistical sample are observed 
between our results and those of Na et al. [\ref{na_han}]. 

\subsection{Particle-laden turbulent channel flow with heat transfer and
momentum/energy two way coupling}
\label{sec3c}

The main purpose of this section is to analyze the modifications on 
the flow field due to the mutual interaction between fluid and particles.
In particular, results obtained from two-way coupling simulations at $Pr=3$
are compared against those obtained from the corresponding one-way coupling
simulations, in which particle feedback
on the flow is neglected: $f_{2w}=0$ and $q_{2w}=0$ in Eqns. (\ref{ns_adim}) and (\ref{en_adim}).
We remark here that, even though the problem of fluid-particle momentum
coupling in two-phase flows has been widely investigated [\ref{pb96},\ref{Boivin},\ref{Kiger}],
much less effort has been devoted to the energy coupling problem.
Currently, numerical studies on this problem are available
for homogeneous shear flow [\ref{shotorban}] and 
for homogeneous isotropic turbulence [\ref{jaberi}].
To our knowledge, this is the first attempt to study (by means of DNS) 
both momentum and energy
coupling between fluid and particles in wall-bounded flow.

\subsubsection{Velocity field modifications by particles}

\begin{figure}[t]
\centerline{\hbox{\includegraphics[width=6.5cm,angle=270]{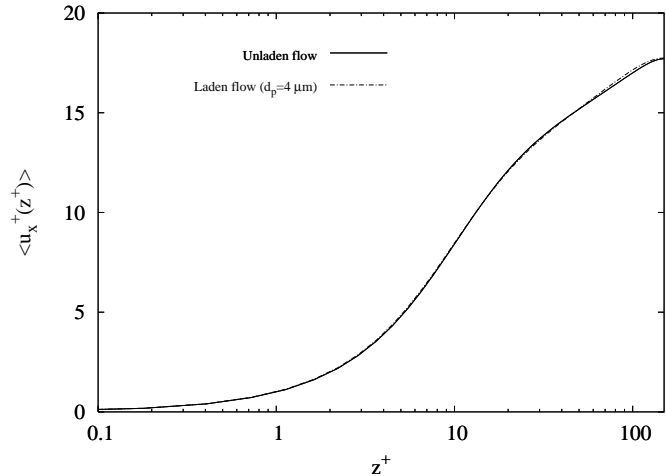}}}
\vspace{-0.2cm}
\caption{ Mean streamwise fluid velocity, $<u_x^+(z^+)>$, at $Pr=3$: comparison between
unladen flow (no momentum/energy coupling) and flow laden with $d_p=4~\mu m$
particles (with momentum/energy coupling).
}
\label{u_mean2w}
\end{figure}

The effect of particles on the mean fluid
velocity, $<u_x^+(z^+)>$, is shown in Fig. (\ref{u_mean2w}).
The solid line refers to the simulation without particles,
while the dot-dashed line refers to the two-way coupling
simulation with the $4~\mu m$ particles.
Profiles are averaged in space
(over the streamwise and spanwise directions) and time 
(over a time span of 180 wall units).
As expected, mean velocity profiles, normalized here
by the unladen flow shear velocity $u^*$,
deviate only slightly (if not negligibly, as in the viscous
sublayer) from each other.
A careful examination of Fig. (\ref{u_mean2w}) indicates that the
effect of particles is to shift the velocity profiles slightly
toward smaller values in the buffer region
($5<z^+<30$) and toward higher values in the outer region
($z^+>30$).
Comparison of the mean velocity profiles for the $8~\mu m$
particle case (not shown here for brevity) indicate
no observable effect.

\begin{figure}[h]
\centerline{\hbox{\includegraphics[width=6.4cm,angle=270]{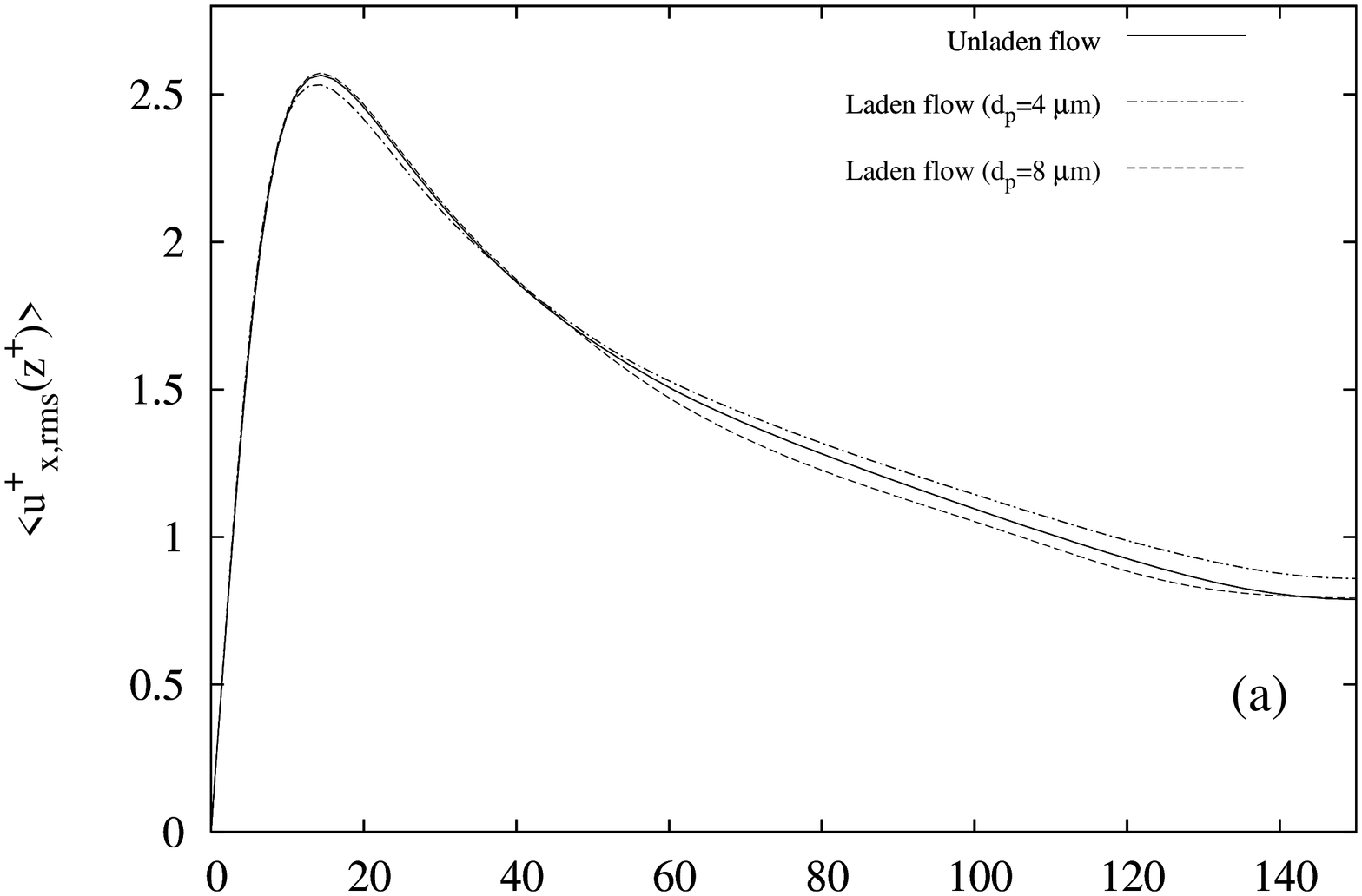}}}
\vspace{-0.72cm}
\centerline{\hbox{\includegraphics[width=6.4cm,angle=270]{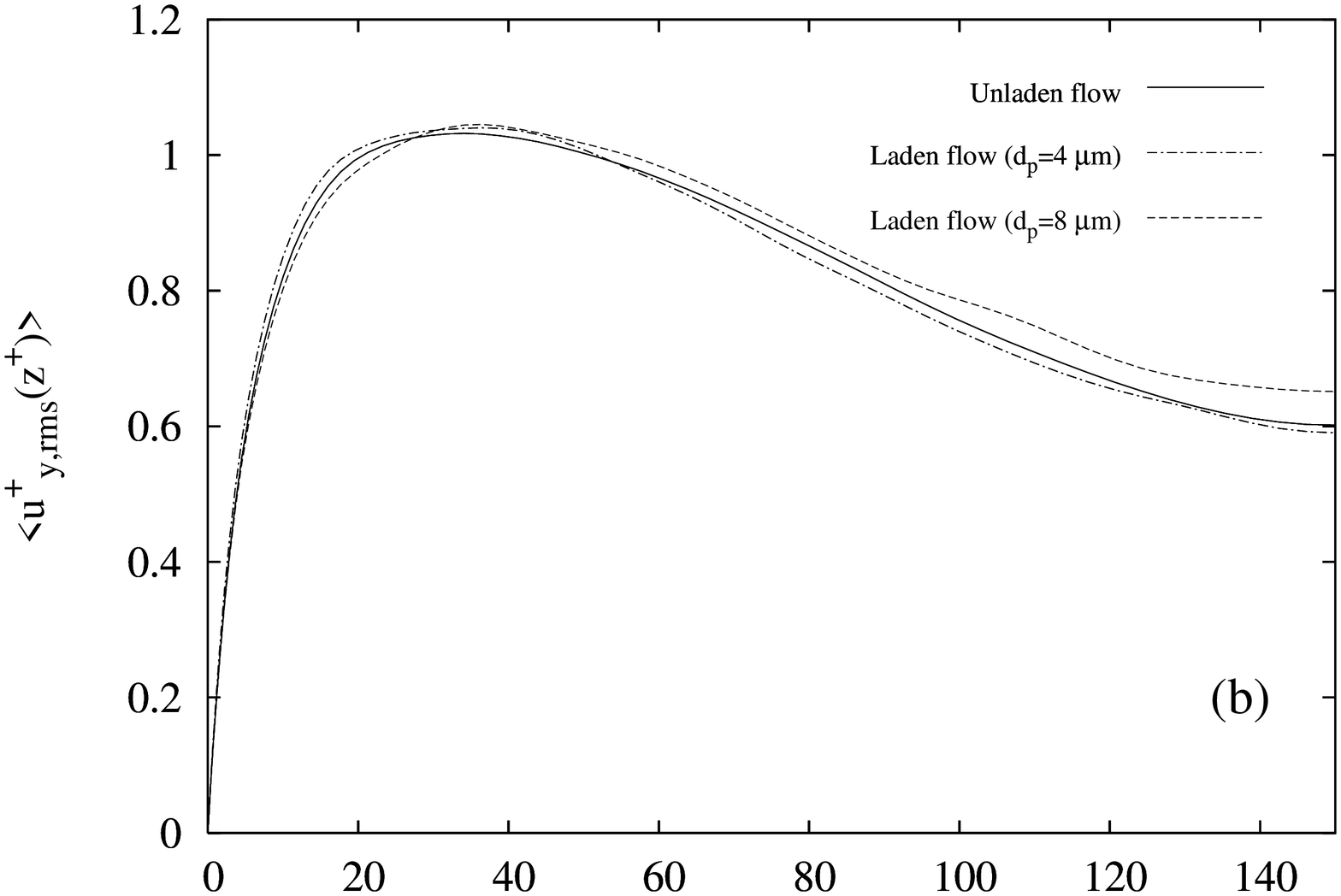}}}
\vspace{-0.72cm}
\centerline{\hbox{\includegraphics[width=6.4cm,angle=270]{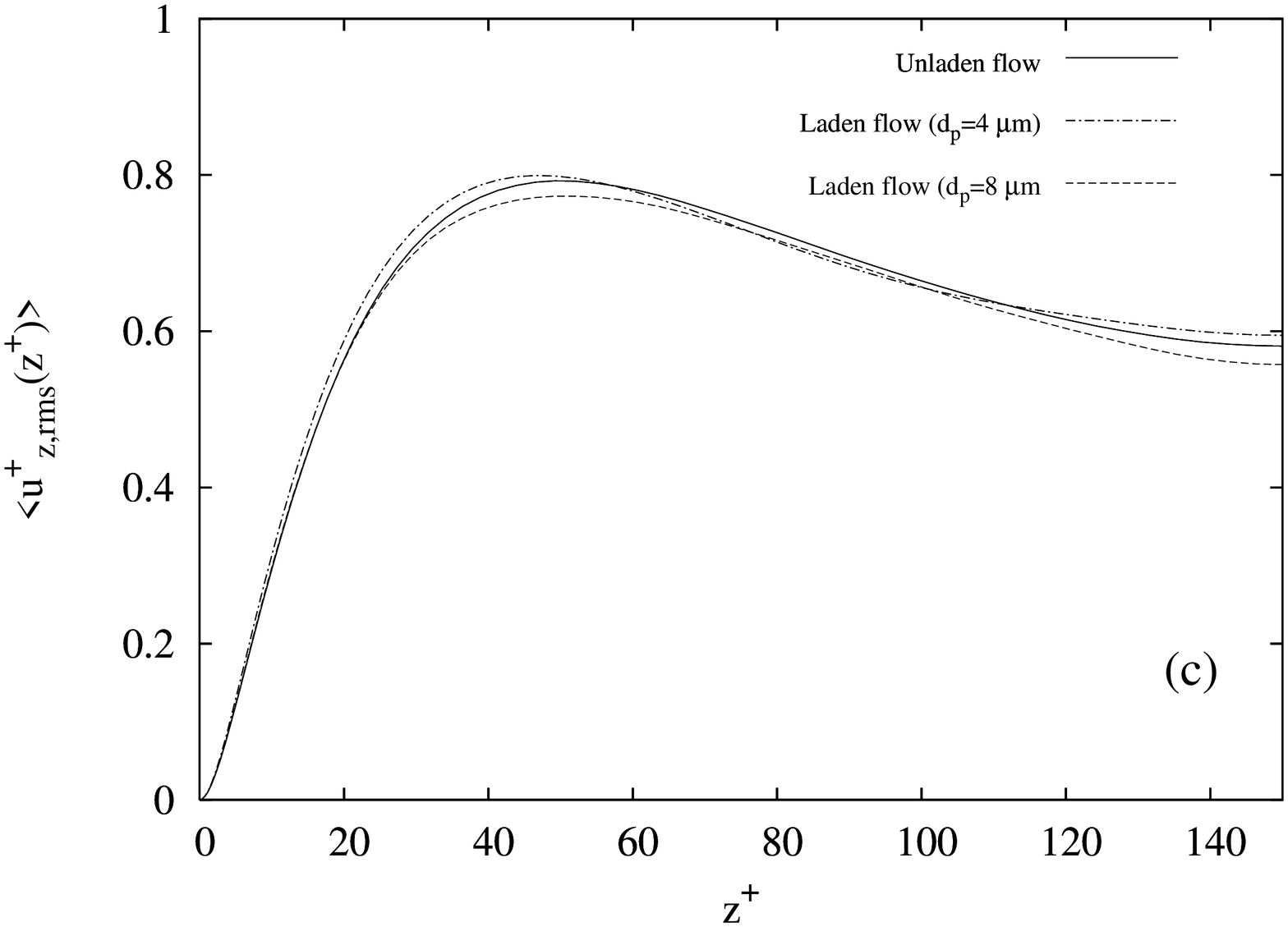}}}
\vspace{-0.1cm}
\caption{ Rms of mean fluid velocity components, $<u^+_{i,rms}(z^+)>$, at $Pr=3$:
comparison between unladen flow (no momentum/energy
coupling) and flow laden with $d_p=4$ $\mu m$ particles and with $d_p=8$ $\mu m$ particles
(with momentum/energy coupling). Panels: (a) streamwise rms, $<u^+_{x,rms}(z^+)>$;
(b) spanwise rms, $<u^+_{y,rms}(z^+)>$;
(c) wall-normal rms, $<u^+_{i,rms}(z^+)>$.}
\label{u_rms2w}
\end{figure}

The behavior of the turbulence intensities,
given by the rms of the fluid velocity fluctuations, $<u^+_{i,rms}(z^+)>$,
shows larger differences as presented in Fig. (\ref{u_rms2w}).
Again, the solid lines refer to the simulation without particles,
whereas the dot-dashed and the dashed lines refer to the two-way coupling
simulation with the $4~\mu m$ particles and with the $8~\mu m$ particles,
respectively.
It appears that particles do not affect much the intensities in the
near-wall region but do substantially change them in the buffer
region, particularly where the profiles develop a peak,
and at the channel centerline.
For each rms component, local changes of opposite sign
depending on particle inertia are observed with respect
to the reference unladen-flow values.

\subsubsection{Temperature field modifications by particles}

\begin{figure}[t]
\centerline{\hbox{\includegraphics[width=6.4cm,angle=270]{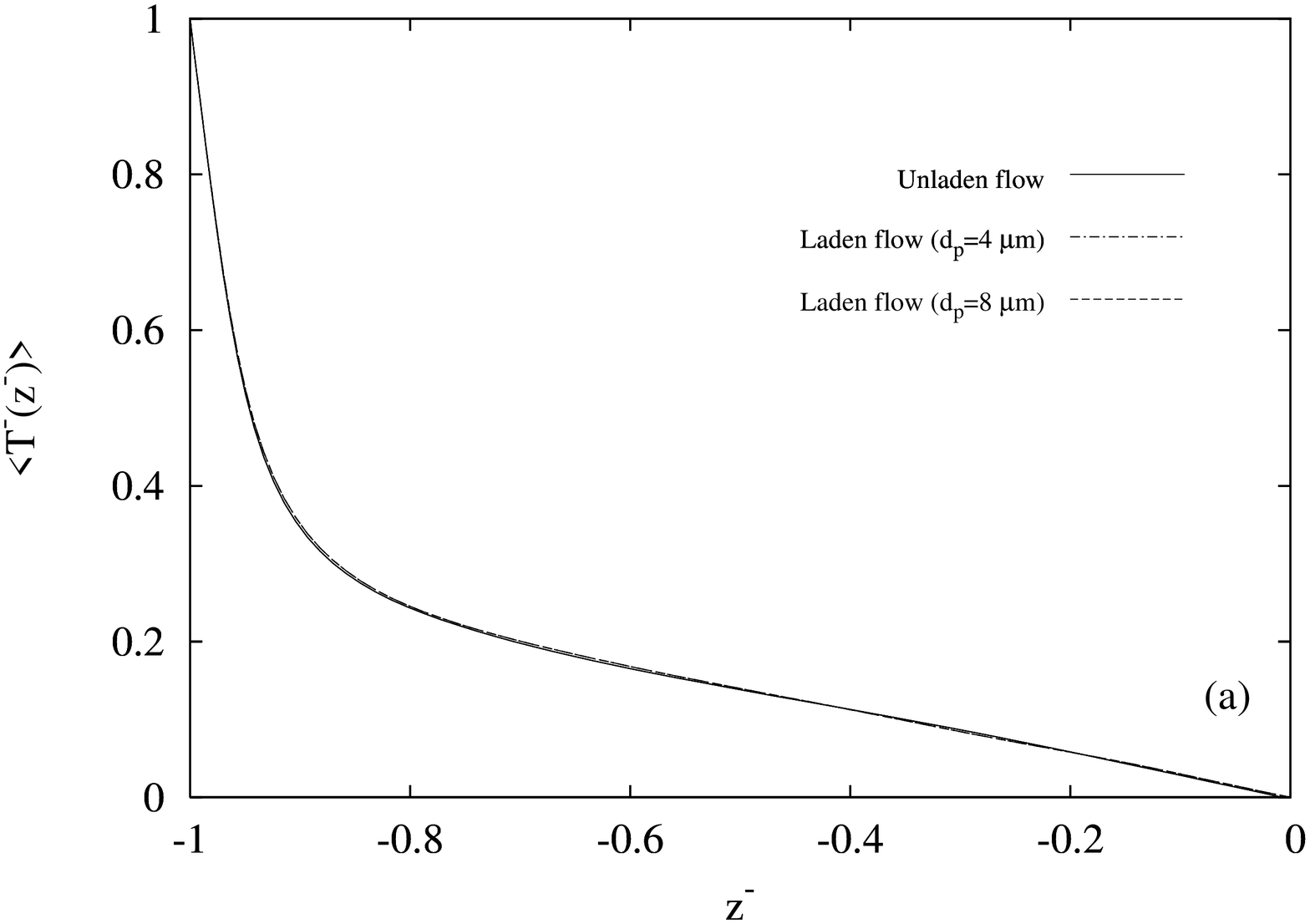}}}
\centerline{\hbox{\includegraphics[width=6.4cm,angle=270]{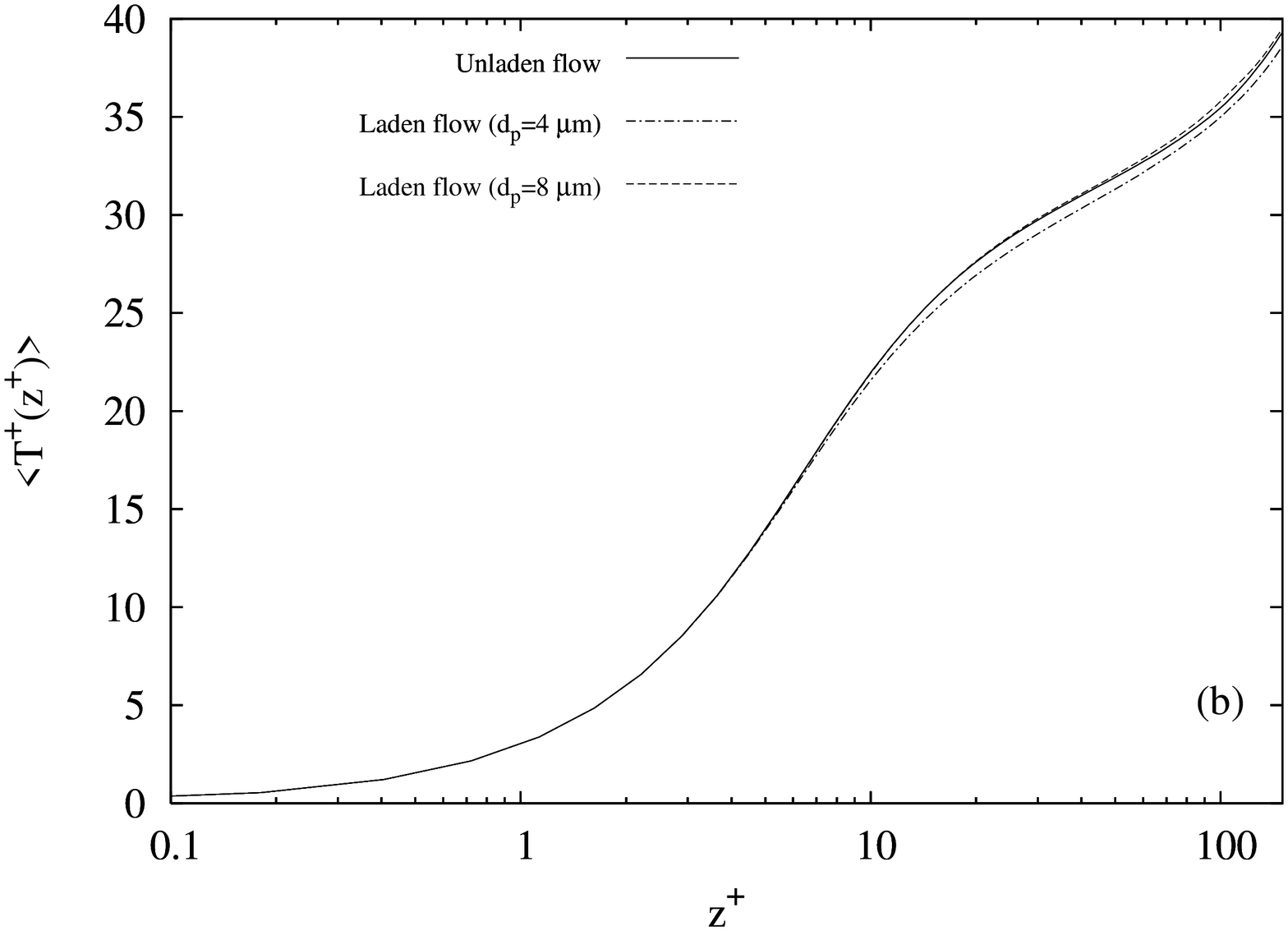}}}
\vspace{-0.2cm}
\caption{ Mean fluid temperature at $Pr=3$: comparison between unladen flow (no momentum/energy
           coupling) and flow laden flow with $d_p=4$ $\mu m$ particles and with $d_p=8$ $\mu m$ particles
(with momentum/energy coupling). Panels: (a) $<T^-(z^-)>$ (lin-lin scale), (b) $<T^+(z^+)>$ (log-lin scale).}
\label{t_mean2w}
\end{figure}

Fig. (\ref{t_mean2w}) shows the mean fluid temperature profiles
in inner units, $<T^-(z^-)>$, (Fig. \ref{t_mean2w}a) and in
wall units, $<T^+(z^+)>$ (Fig. \ref{t_mean2w}b).
Lines are as in Fig. (\ref{u_mean2w}).
Visual inspection of Fig. (\ref{t_mean2w}a) does not reveal
significant differences in the profiles. However, computing
the local value of the wall-normal temperature gradient,
$\d$<$T(z)$>$/\d z$,
right at the wall, we obtain an increase
of
roughly $2~\%$
~\\
\vspace{-0.8cm}

for the two-way coupling
simulation with the $4~\mu m$ particles and a decrease of
$0.2~\%$ for the two-way coupling
simulation with the $8~\mu m$ particles.
This finding is important because changes produced
by the particles to the wall-normal temperature gradient
directly reflect upon the
heat flux at the wall, $q_w$, through the following expression:
\begin{equation}
q_{w}=\frac{1}{Pr}\frac{\d <T(z)>}{\d z}~.
\end{equation}
In turn, a change in the value of $q_w$ will eventually correspond to a
change in the value of the total turbulent
heat flux $q_{z,tot}$, defined as:
\begin{equation}
q_{z,tot}=q_{w}-<T'w'>~,
\end{equation}
under fully developed conditions.
Besides being perhaps the most significant
result of the present paper from a quantitative viewpoint,
the increase of removable heat flux at the wall for the
smaller $4~\mu m$ particles case and the decrease of removable heat flux
for the larger $8~\mu m$ particles case have also an effect on the
friction temperature, $T^*$. According to its definition (see Sec.
\ref{unladen}), the friction temperature computed in the two-way coupling
simulations with the $4~\mu m$ particles will increase with respect to
the unladen flow case, whereas it will decrease in the simulations with
the $8~\mu m$ particles.
As a consequence, starting from the reference profile relative to
the unladen flow case, the mean fluid temperature profile computed
for the $4~\mu m$ particle case will shift toward smaller values
(particularly outside the viscous sublayer) whereas it will shift
toward slightly higher values when the larger $8~\mu m$ particles
are considered, as apparent from Fig. (\ref{t_mean2w}b).

\begin{figure}[t]
\centerline{\hbox{\includegraphics[width=6.5cm,angle=270]{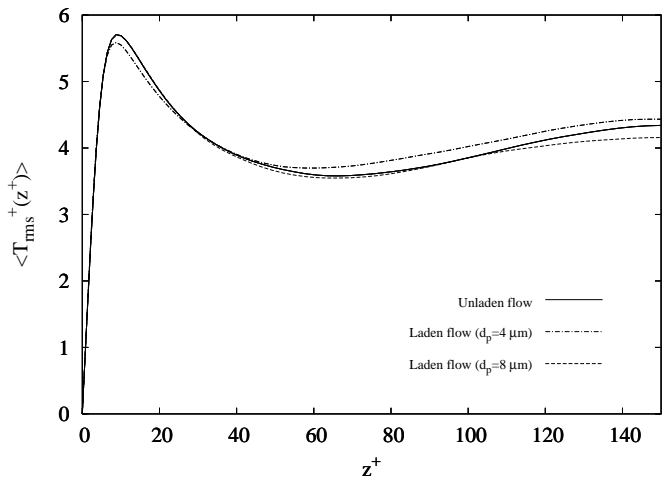}}}
\vspace{-0.2cm}
\caption{ Rms of fluid temperature, $<T^+_{rms}(z^+)>$, at $Pr=3$: comparison between unladen flow (no momentum/energy
           coupling) and flow laden with $d_p=4$ $\mu m$ particles and with $d_p=8$ $\mu m$ particles
 (with momentum/energy coupling). }
\label{t_rms_mean2w}
\end{figure}

Finally, the rms of the fluid temperature fluctuations, $<T^+_{rms}(z^+)>$,
are shown in Fig. (\ref{t_rms_mean2w}). The $4~\mu m$ particles produce a slight
decrease in the peak value of $<T^+_{rms}(z^+)>$ and an increase outside
the buffer region; smaller modifications (mostly limited to the core
region of the flow) are produced by the larger $8~\mu m$ particles.

\subsubsection{Influence of particle inertia and of particle thermal inertia}

The interactions between particles, turbulent momentum transport and 
turbulent heat transport are influenced by the particle response times.
In this section we analyze the Eulerian statistics of the particles
in comparison with those of the fluid. 
Specifically, particles will acquire and lose momentum and heat
at a rate proportional to the inverse of their response times,
$\tau_p$ and $\tau_T$ respectively.

\begin{figure}[t]
\centerline{\hbox{\includegraphics[width=6.5cm,angle=270]{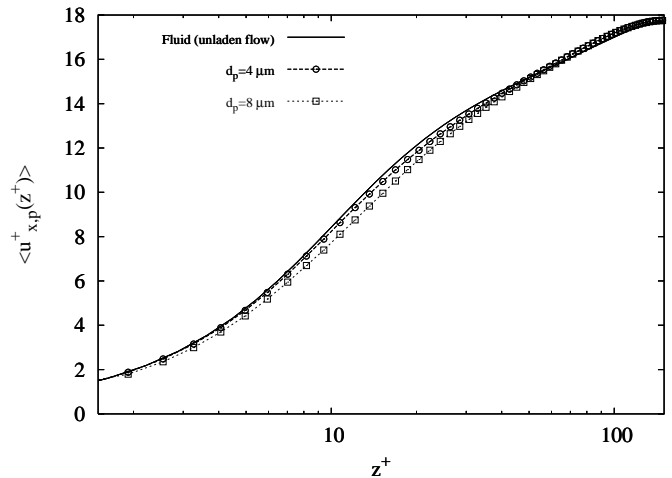}}}
\vspace{-0.2cm}
\caption{Mean particle streamwise velocity, $<u^+_{x,p}(z^+)>$, at $Pr=3$. Symbols: ($\bigcirc$) $d_p=4$ $\mu m$ particles,
($\square$) $d_p=8$ $\mu m$ particles. The mean fluid streamwise velocity profile relative to the unladen flow
case (solid line) is also shown for sake of comparison.
}
\label{umean_part_set1}
\end{figure}

\begin{figure}[h]
\centerline{\hbox{\includegraphics[width=6.5cm,angle=270]{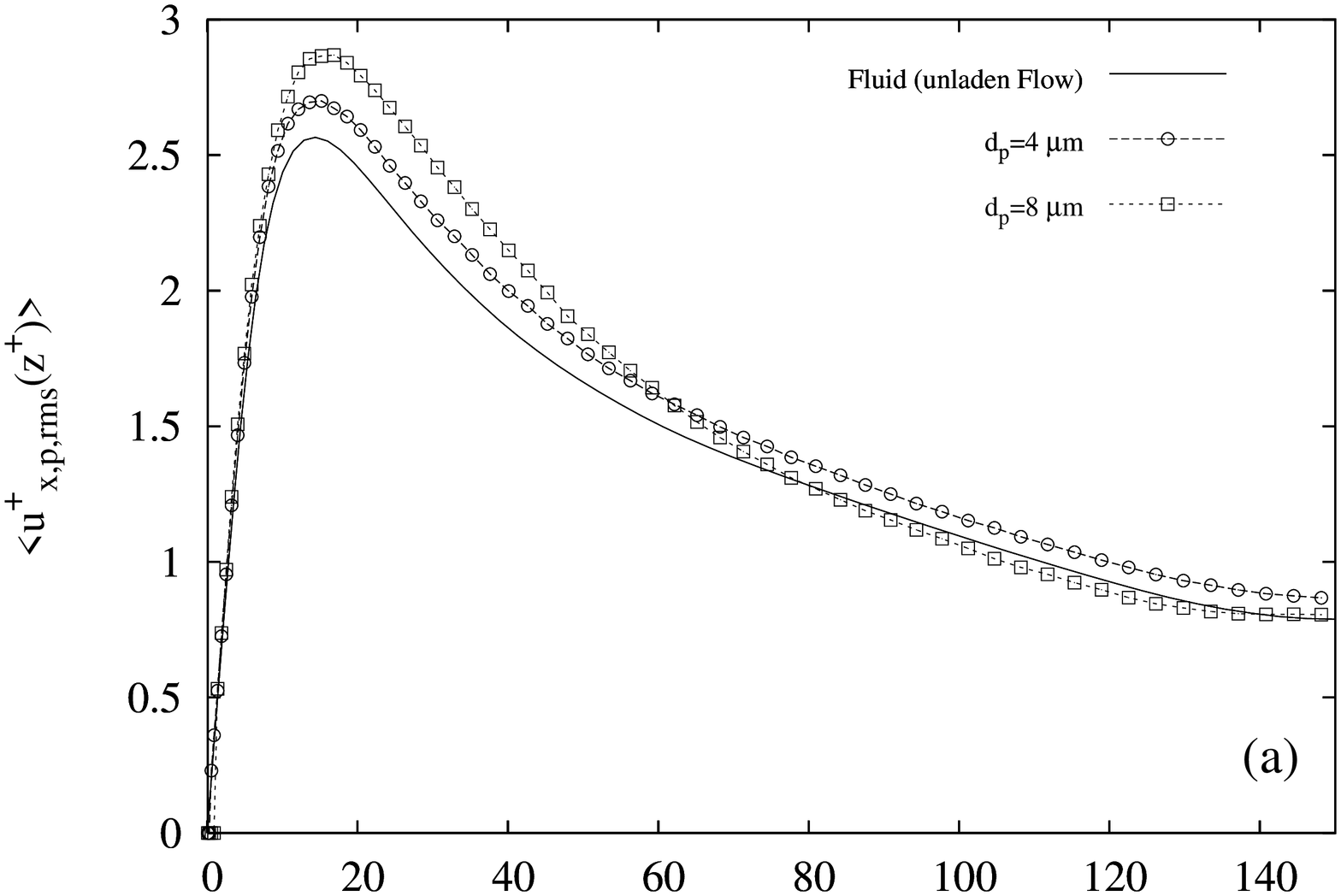}}}
\vspace{-0.72cm}
\centerline{\hbox{\includegraphics[width=6.5cm,angle=270]{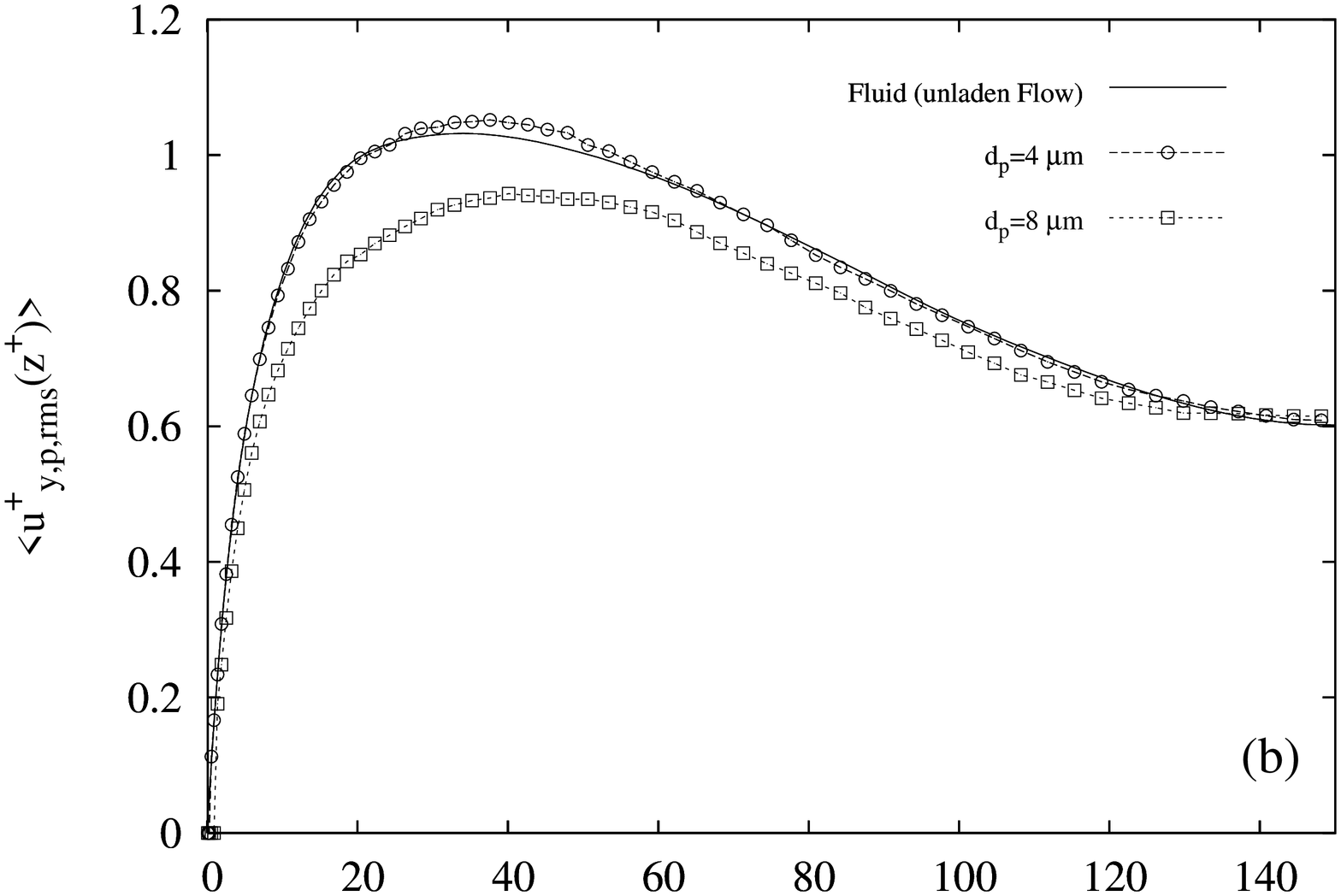}}}
\vspace{-0.72cm}
\centerline{\hbox{\includegraphics[width=6.5cm,angle=270]{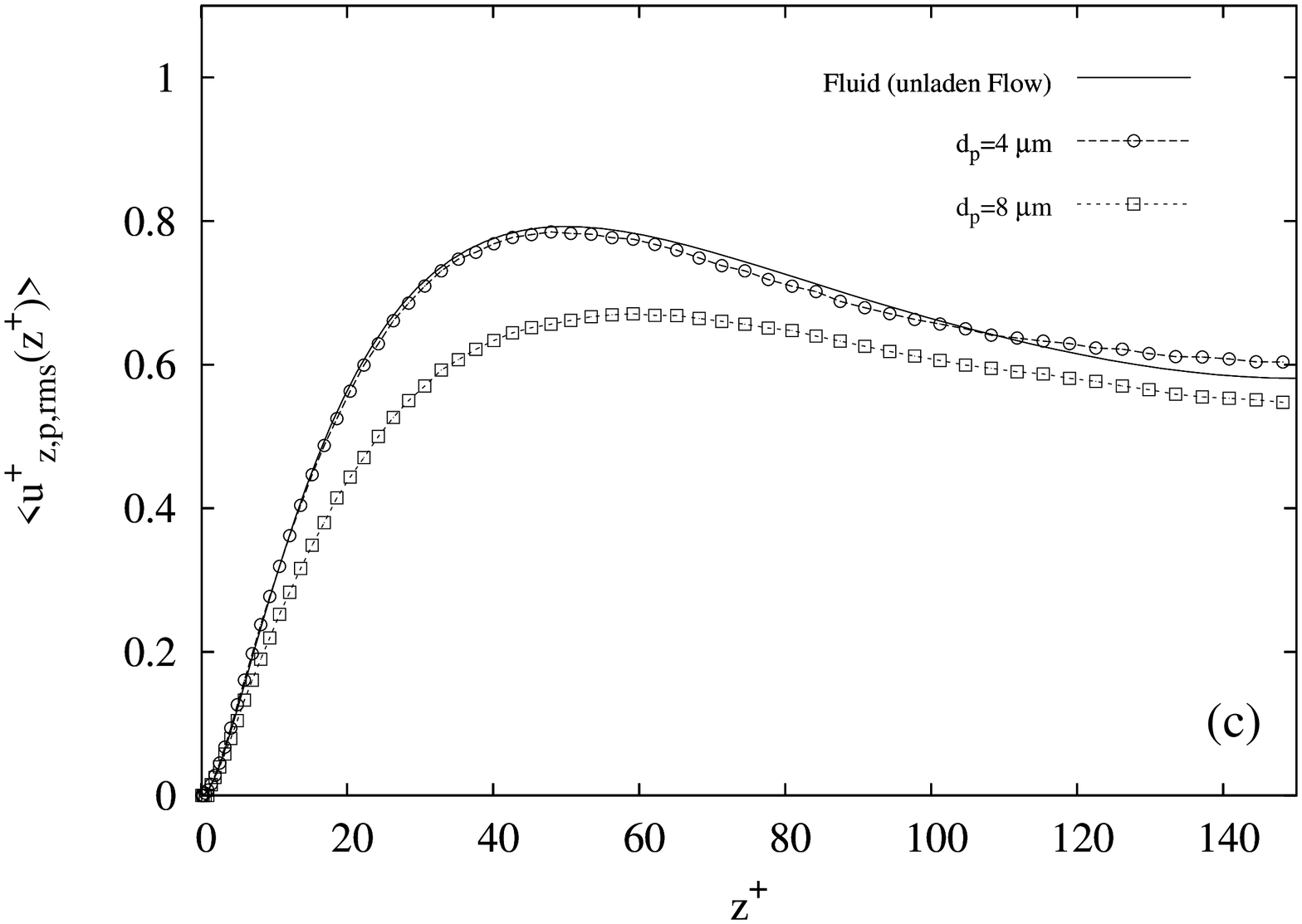}}}
\vspace{-0.28cm}
\caption{Rms of particle velocity components at $Pr=3$.
Symbols: ($\bigcirc$) $4~\mu m$ particles,
($\square$) $8~\mu m$ particles.
Panels: (a) streamwise rms, $<u^+_{x,p,rms}(z^+)>$, (b) spanwise rms, $<u^+_{y,p,rms}(z^+)>$,
(c) wall-normal rms, $<u^+_{z,p,rms}(z^+)>$.
The rms of fluid velocities relative to the unladen flow
case (solid line) is also shown for sake of comparison.
}
\label{urms_part_set1}
\end{figure}

Fig. (\ref{umean_part_set1}) shows the mean streamwise velocity profile,
$<u^+_{x,p}(z^+)>$, averaged in time and along the homogeneous
directions, for both
fluid and particles as a function of the wall-normal coordinate,
$z^+$.
Differences are readily visible,
with a consistent evidence of the effect of particle inertia.
Smaller particles ($St=1.5$, $d_p=4~\mu m$) behave more like fluid 
tracers 
and their average velocity profile (circles) almost match
that of the fluid (solid line). 
As the particle response time increases 
($St=6$, $d_p=8~\mu m$), the average particle velocity (squares)
is seen to lag the average fluid velocity, in particular outside the
viscous sublayer.  
Present results agree well with those 
reported by van Haarlem et al. [\ref{haarlem}] and by Portela et al. [\ref{portela}],
who however used a one-way approach.
This shows that, for the current size, density and overall concentration
of the particles, 
modifications to the mean streamwise velocity profile
due to momentum and energy coupling appear negligible.
As already observed by van
Haarlem et al. [\ref{haarlem}], inertial particles dispersed in a turbulent
flow do not sample the flow field homogeneously and tend 
to avoid areas of high vorticity preferring areas characterized by
lower-than-mean streamwise fluid velocity and by high strain rate.
This gives the characteristic velocity lag in the region $5<z^+<50$.
This effect is confirmed by data shown in Fig. (\ref{urms_part_set1}a),
where the root mean square of particles streamwise velocity (circles and squares)
is compared to that of the fluid (solid line). 
Fluctuations of particle streamwise
velocity are larger than those of the fluid, this difference
becoming more evident as particle response time
increases [\ref{portela}].
From a physical viewpoint, the difference in the
streamwise values suggests that the gradients in the
mean fluid velocity can produce significant fluctuations
of the streamwise particle velocity. 
This effect seems more pronounced in the case of heavy particles, with
larger Stokes number and a longer ``memory''.
An opposite behavior is observed in the spanwise
direction and in the wall-normal 
direction (Figs. (\ref{urms_part_set1}b) and (\ref{urms_part_set1}c), respectively),
where the fluid velocity field
has zero mean gradient. The turbulence intensity of the
particles is very close to that of the fluid for 
particles with small inertia ($St=1.5$, circles) and lower than that 
of the fluid for particles with higher inertia ($St=6$, squares).
This is mainly due 
to two mechanisms acting in tandem. The first mechanism is
preferential concentration of particles in low-speed regions,
characterized by lower turbulence intensity [\ref{s05}]. The
second is the filtering of high frequencies or wavenumber
fluctuations done by particles due to their inertia.
The inertial filtering damps turbulence intensity of the particles
in the wall-normal direction and in the spanwise
direction. Similar filtering effects have been observed in
homogeneous turbulence [\ref{reeks}].

\begin{figure}[t]
\centerline{\hbox{\includegraphics[width=6.4cm,angle=270]{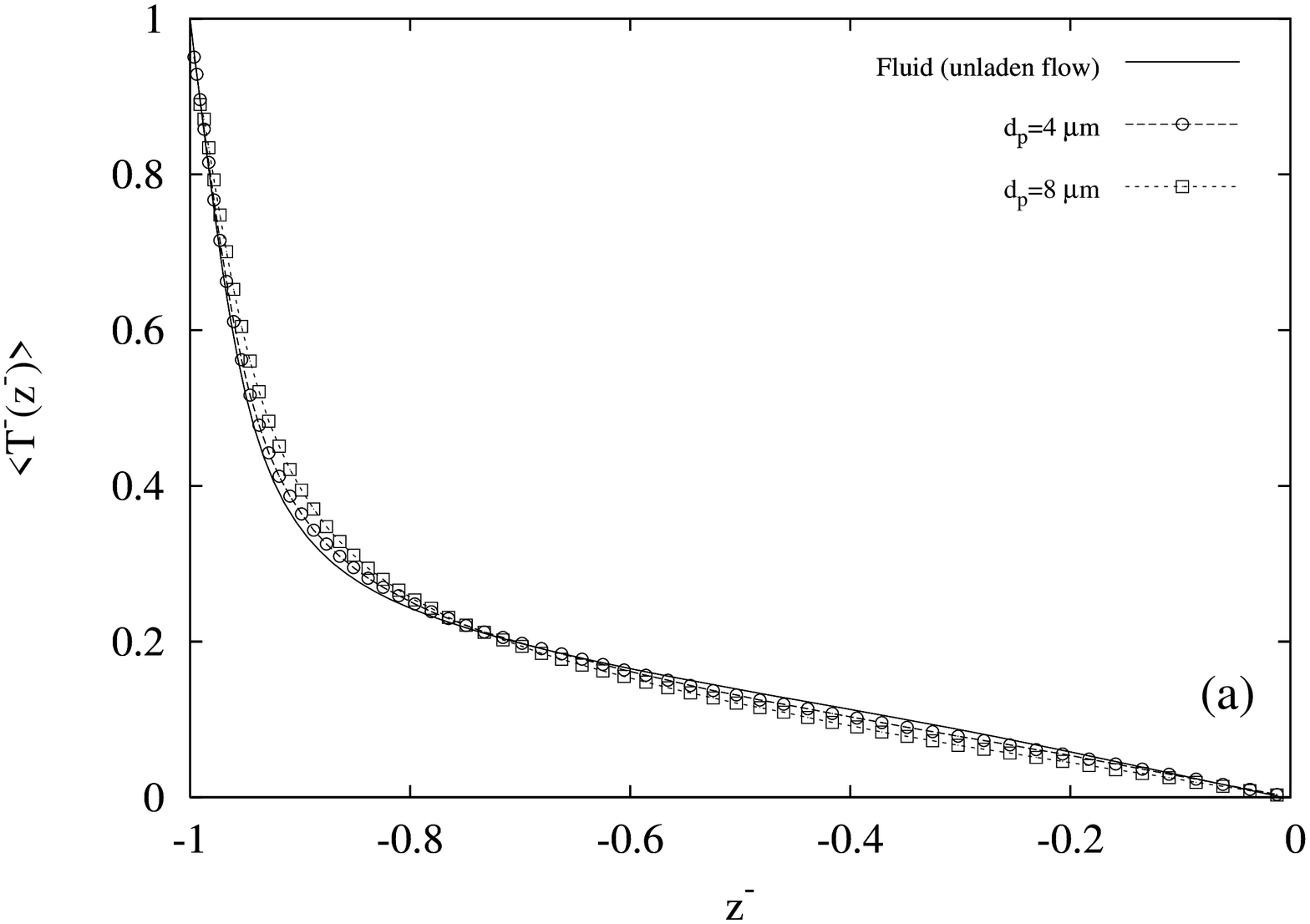}}}
\centerline{\hbox{\includegraphics[width=6.4cm,angle=270]{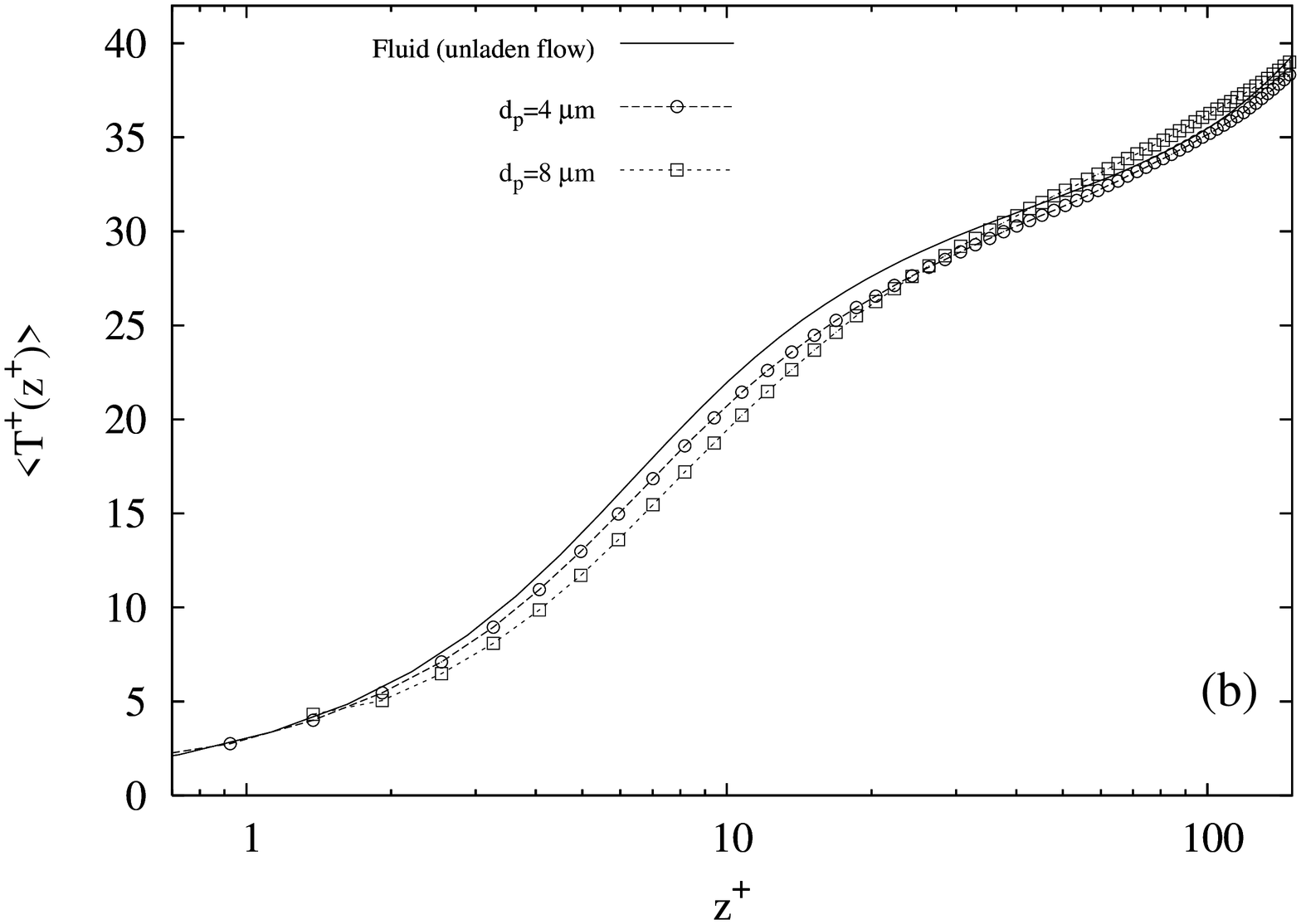}}}
\vspace{-0.2cm}
\caption{ Mean particle temperature at $Pr=3$. Symbols: ($\bigcirc$) $4~\mu m$ particles,
($\square$) $8~\mu m$ particles.
Panels: (a) $<T^-_p(z^-)>$ (linear-linear scale), (b) $<T^+_p(z^+)>$ (log-linear scale).
The mean fluid velocity relative to the unladen flow
case (solid line) is also shown for sake of comparison.}
\label{tmean_part_set1}
\end{figure}

\begin{figure}[t]
\centerline{\hbox{\includegraphics[width=6.5cm,angle=270]{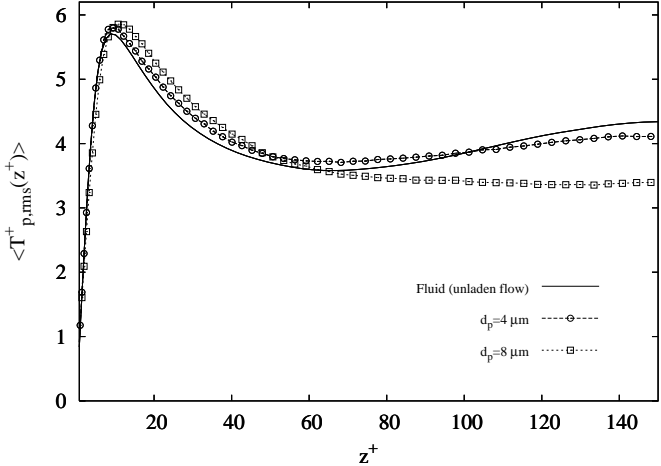}}}
\vspace{-0.2cm}
\caption{ Rms of particle temperature, $<T^+_{p,rms}(z^+)>$, at $Pr=3$. Symbols: ($\bigcirc$) $4~\mu m$ particles,
($\square$) $8~\mu m$ particles.
The rms of fluid temperature relative to the unladen flow
case (solid line) is also shown for sake of comparison.}
\label{trms_part_set1}
\end{figure}

The statistical behavior of the thermal field for the dispersed phase 
is presented in Figs. (\ref{tmean_part_set1}) and (\ref{trms_part_set1}).
Fig. (\ref{tmean_part_set1})
shows the mean temperature 
profiles
for particles and fluid as a function of the wall-normal coordinate.
In Fig. (\ref{tmean_part_set1}a) variables are expressed in outer units,
whereas in Fig. (\ref{tmean_part_set1}b) variables are shown in wall units and in semi-logarithmic scale.
The solid line refers to the average unladen fluid 
temperature, while circles and squares 
refer respectively to the simulations with the $4~\mu m$ particles and with the $8~\mu m$
particles.
Fig. (\ref{tmean_part_set1}a) indicates that, for both particle sets,
the particle temperature in the near-wall region ($0<z^+<30$) is
higher than that of the fluid, while it 
reaches lower values in the outer region ($30<z^+<150$).
Particle temperature higher than that of the  fluid is probably due to the deposition/resuspension 
mechanisms, which bring particles to the near-wall region 
(where they are characterized by high temperature value differences) and drive them 
toward the core region [\ref{ms02},\ref{s05}].
When a particle initially close to the wall 
is entrained toward the core region, it will adapt its temperature at a rate depending
on particle thermal inertia, thus maintaining a temperature higher than that of the fluid.
In the outer region, the perspective is overturned, and particles are characterized 
by a temperature lower than that of the fluid.
The effect of  particle thermal inertia is also visible by observing the rms
of particle temperature profiles, $<T^+_{p,rms}(z^+)>$, shown in Fig. (\ref{trms_part_set1}). 
These profiles are qualitatively similar to those for the streamwise rms component 
(Fig. \ref{urms_part_set1}a),
due to the role played by particle thermal inertia in transferring heat, similar to that played
by particle inertia in transferring momentum.
Although a qualitative similarity between the temperature and the streamwise velocity
rms can be observed, quantitative differences are visible. In particular, the peak of the 
streamwise velocity rms exhibits values higher than the temperature rms.
These quantitative differences are due to the fact that particle inertia is
larger than particle thermal inertia and causes streamwise velocity fluctuations
higher than thermal fluctuations.

\subsubsection{Instantaneous features of turbulent transfer of momentum, heat and particles}

A number of DNS-based works have been successful in clarifying the role of the instantaneous
realizations of the Reynolds stresses in transferring momentum [\ref{adrian}],
heat [\ref{na_han}] and particles [\ref{ms02}]. We refer to these works and to
the references therein for a more comprehensive description of the phenomena.
It suffices here to show on a qualitative basis some relations among
momentum, heat and particles transfer mechanisms.

\begin{figure}[t]
\vspace{-1.0cm}
\centerline{\includegraphics[width=6.5cm,angle=270]{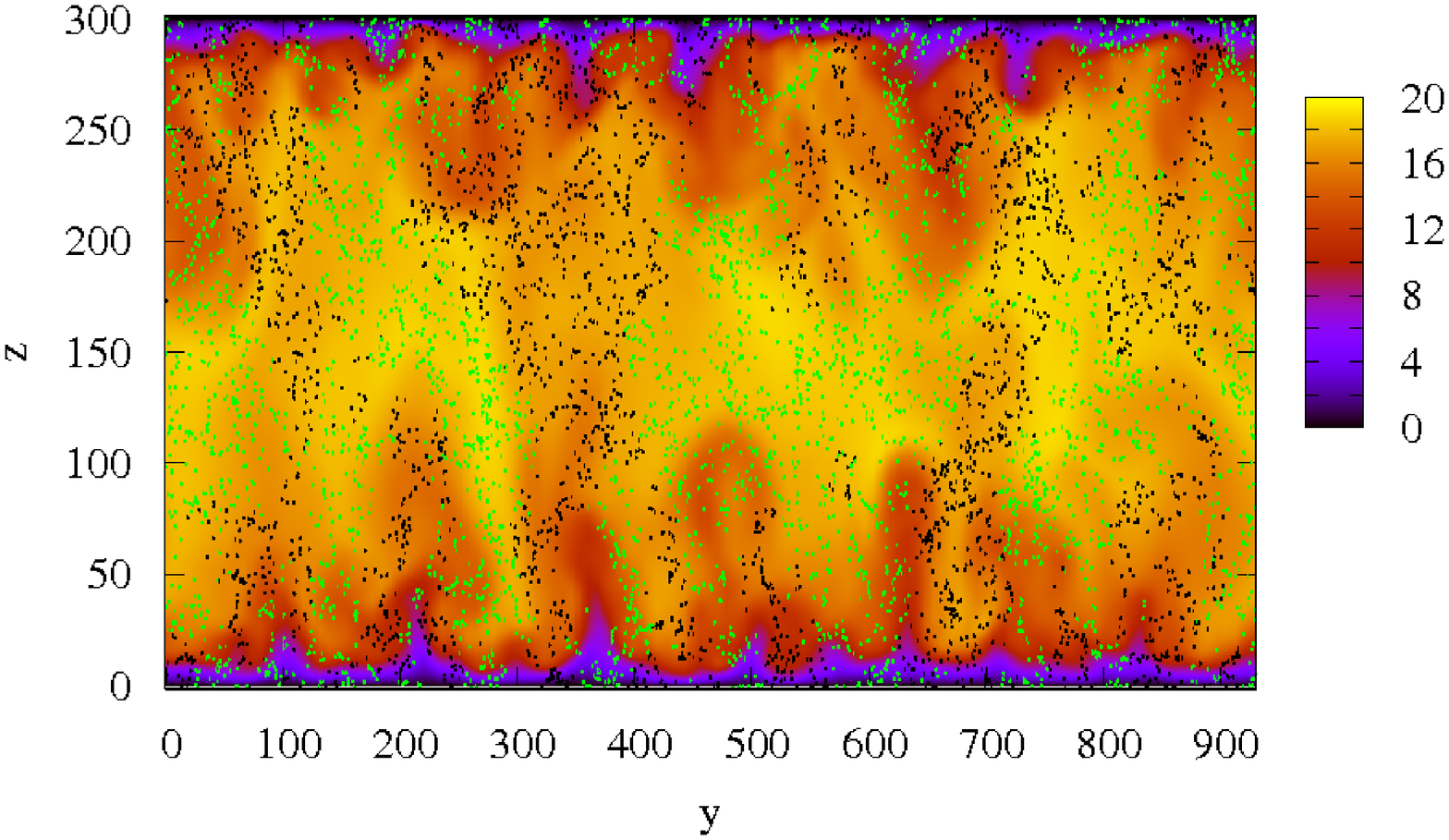}}
\centerline{\includegraphics[width=8.3cm,angle=270]{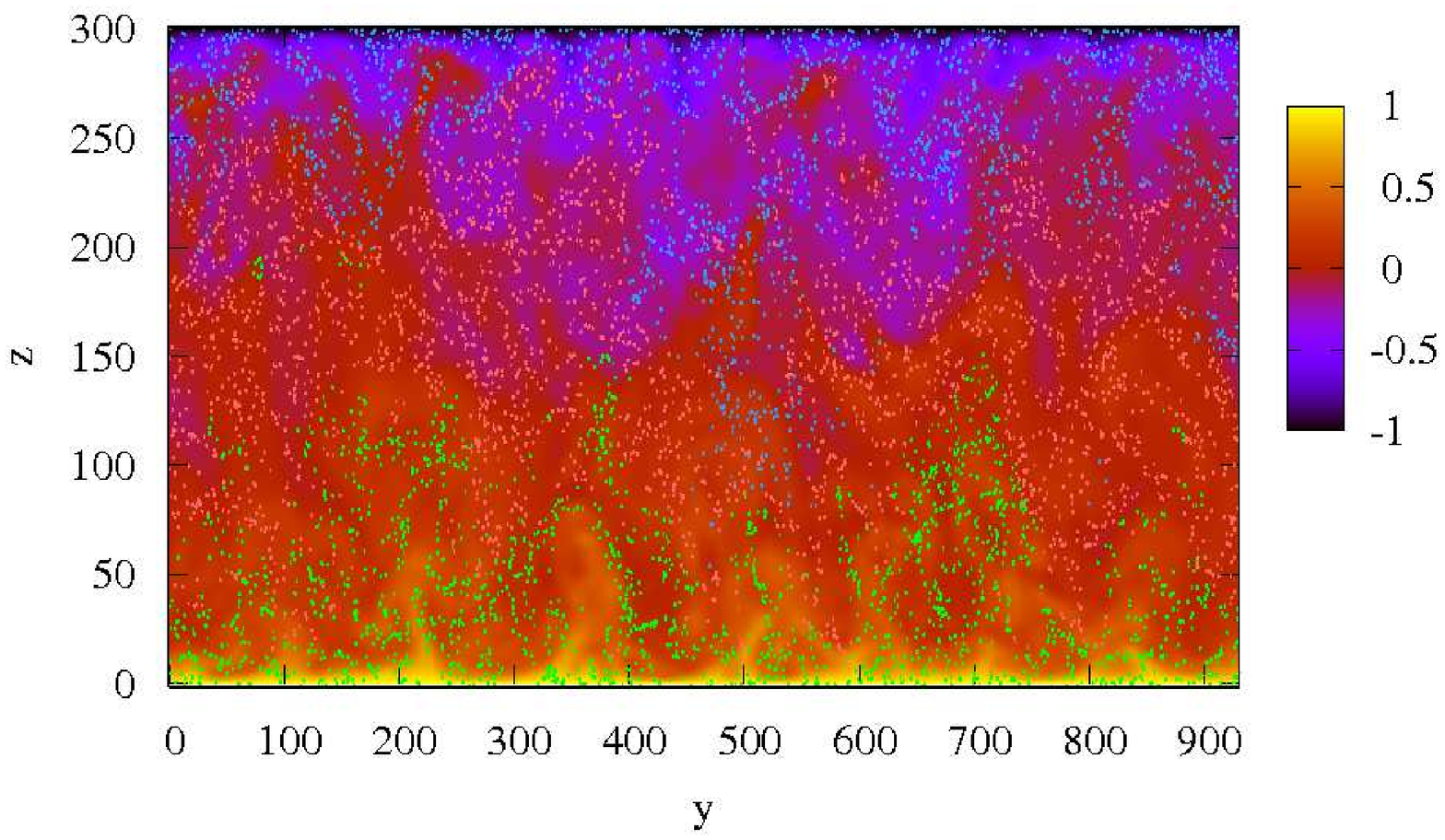}}
\vspace{-4.2cm}
\caption{ Front view of particle instantaneous distribution superimposed to the
fluid velocity field (top) and to the fluid temperature field (bottom)
at $Pr=3$.}
\label{vect_partn}
\end{figure}

To this aim, we complement the statistical analysis of the previous
sections by showing the complex interactions among
velocity field, temperature field and particles distribution.
In particular, we examine simulation $R3$ ($Re^*=150$, $Pr=3$ and $d_p=4~\mu m$ particles).
We consider a cross section of the flow field, perpendicular to the mean velocity,
cut along a $x-y$ plane.
Consider first Fig. (\ref{vect_partn}a):
color isocontours correspond to
values of the streamwise fluid velocity whereas circles represent particles 
(drawn larger than the real size for visualization purposes) colored by their wall-normal velocity.
In this Figure, green particles have wall-normal velocity directed {\em toward the walls},
while black particles have wall-normal velocity directed {\em away from the walls}.
In this way, just by analyzing  particle wall-normal velocity, it is possible to detect
particle transfer fluxes toward the wall and particle transfer fluxes away from the wall.
It is possible to observe that fluxes of particles are associated with 
fluxes of streamwise momentum (called {\em sweeps} if directed toward the wall and {\em ejections}
if directed away from the wall, as discussed in several previous papers [\ref{ms02},\ref{s05}]).

To draw a link between the heat transfer mechanisms and
the momentum transfer mechanisms we consider also
Fig. (\ref{vect_partn}b), where the particles are shown superimposed to the temperature field for
the same section of Fig. (\ref{vect_partn}a).
In this case color isocontours are
the values of the fluid temperature, whereas circles represent particles colored by their temperature:
specifically, green particles have higher-than-mean temperature,
blue particles have lower-than-mean temperature, and red particles 
have temperature close to the mean.
The Reynolds transport analogy is clearly demonstrated by the behavior of the instantaneous temperature field:
focusing on the wall at $z^+=0$, characterized by higher temperature,
we observe hot fluid plumes raising in correspondence with the raising of
low momentum fluid within an ejection in Fig. (\ref{vect_partn}a).
This just demonstrates the effectiveness of the Reynolds stresses in transporting the fluid close to the wall
towards the center of the channel and viceversa.
Similar observations can be made for the cold wall at $z^+=300$.
Considering now, again in Fig. (\ref{vect_partn}b), the behavior of the particles
in  correlation  with the temperature field, we observe that particles
with higher temperature are ejected from 
the lower wall and are directed towards the upper wall.
The opposite occurs to the colder particles.
In this case, however, it is important to remind that  particle temperature
is not strictly correlated with fluid temperature.
This behavior can be explained by considering that particle trajectories depend on fluid velocity
and particle inertia, but not on fluid or particle temperature.
Particles, driven by the fluid vortices, can thus reach regions characterized by fluid temperature
quite different from particle temperature, causing an 
appreciable heat exchange between the two phases.

\begin{figure}[]
\vspace{-0.1cm}
\centerline{
\hspace{1.3cm}
\includegraphics[width=5.0cm,angle=270]{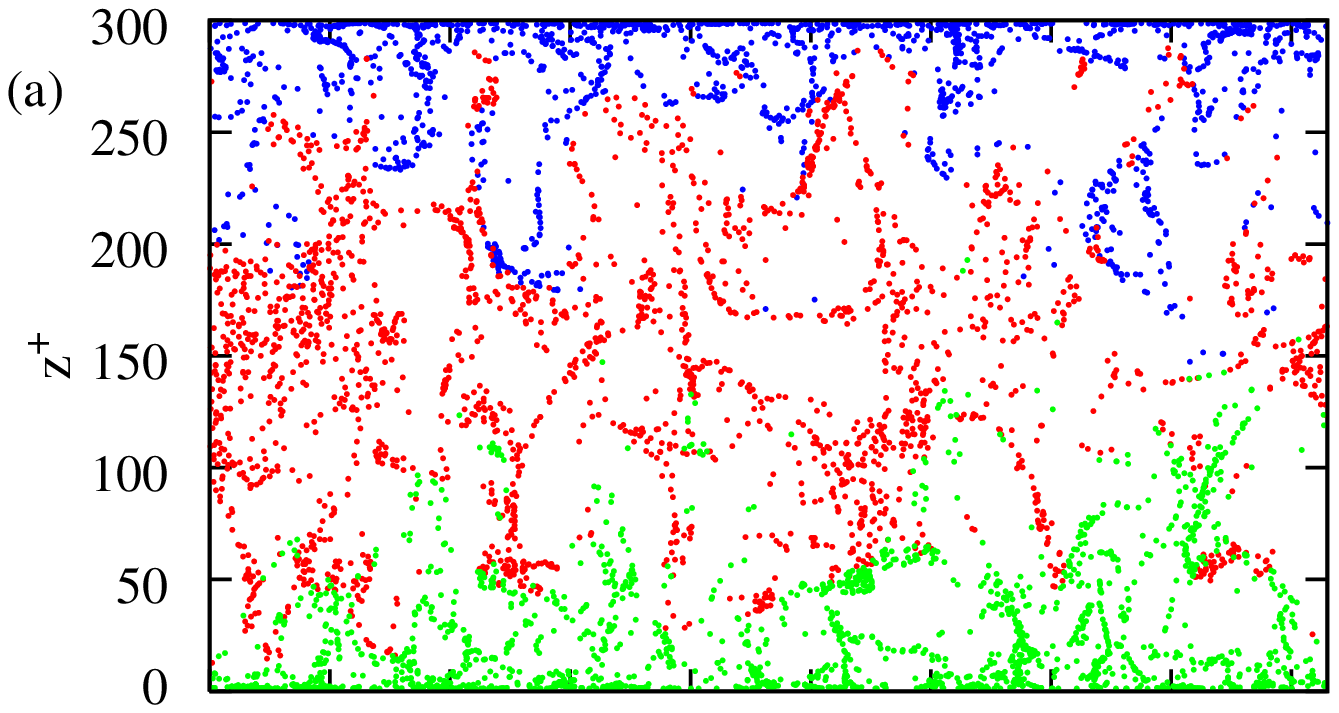}
\includegraphics[height=3.48cm,angle=270]{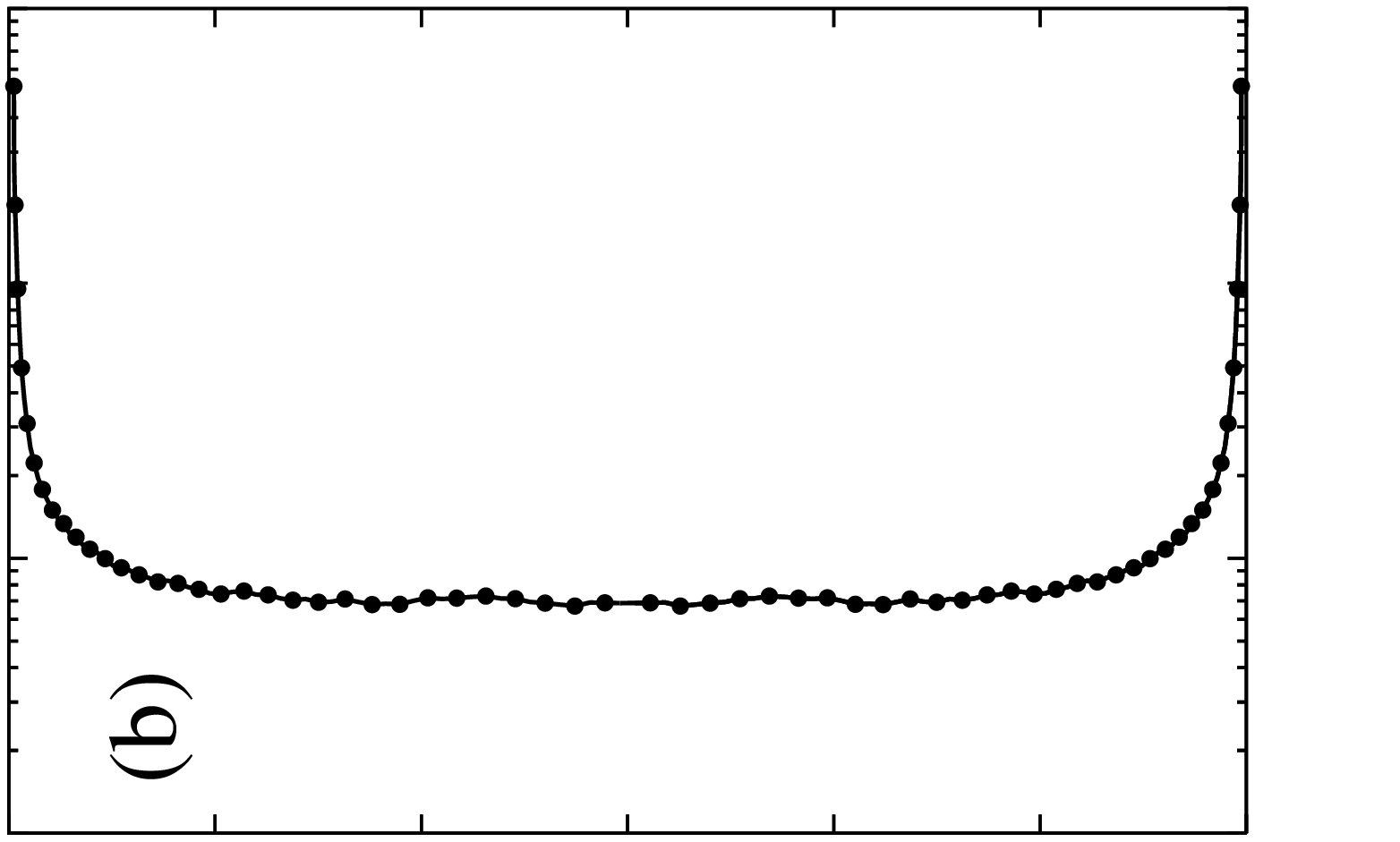}
}
\vspace{-1.4cm}
\centerline{
\hspace{1.3cm}
\includegraphics[width=5.0cm,angle=270]{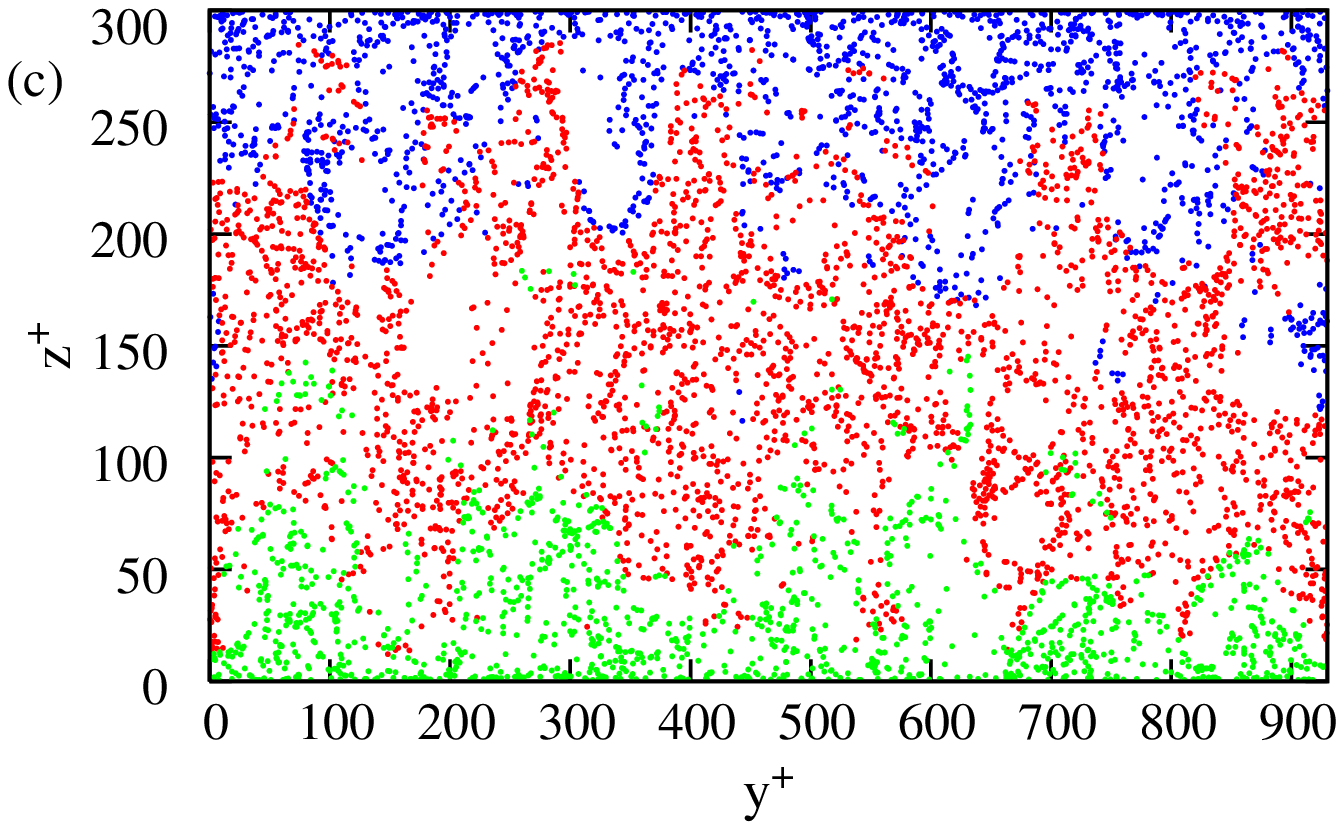}
\includegraphics[height=3.48cm,angle=270]{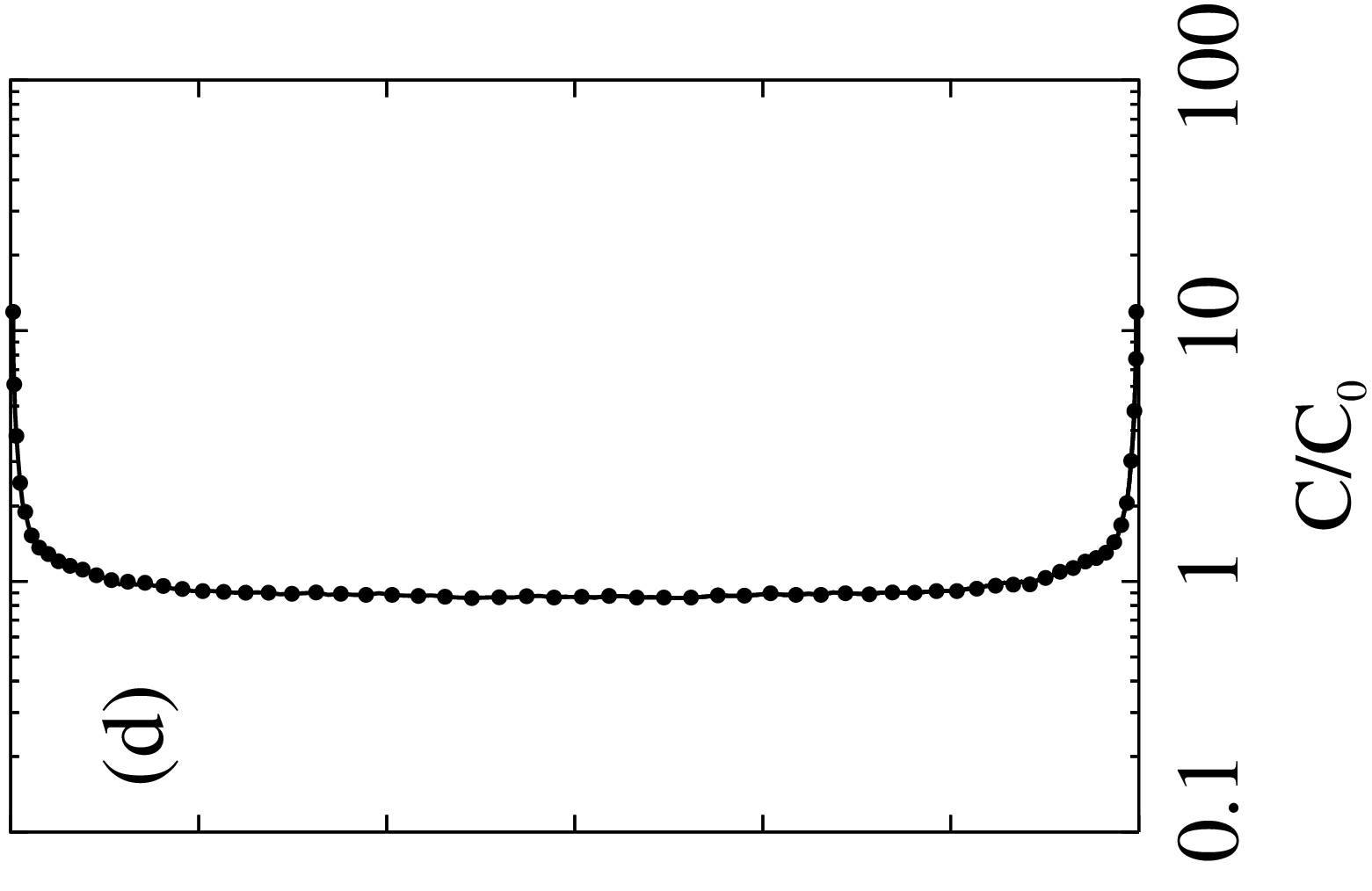}
}
\vspace{-0.5cm}
\caption{ Front view of particle instantaneous distribution (left panels)
and corresponding particle wall-normal concentration, $C/C_0$ (right panels) at $Pr=3$.
Panels: (a-b) $8~\mu m$ particles, (c-d) $4~\mu m$ particles.}
\label{part-conc}
\end{figure}

The efficiency of the overall heat exchange process is conditioned by the degree of non-homogeneity
of particle distribution and by the rate of particle accumulation at the wall.
This observation is qualitatively corroborated by Fig. (\ref{part-conc}), where
the instantaneous cross-sectional distribution (left-hand side panels) and the wall-normal
concentration, $C/C_0$ (right-hand side panels) of both particle sets are compared
at the same time instant.
Note that the $4~\mu m$ particle distribution, shown in Fig. (\ref{part-conc}c),
is the same of Fig. (\ref{vect_partn}), the particle color code being that of Fig.
(\ref{vect_partn}b); particle concentration is computed as particle number
density distribution per unit volume normalized by its initial value [\ref{ms02}].
It is evident that the smaller $4~\mu m$ particles, which produce an increase of heat transfer
at the wall (as discussed in Sec. \ref{sec3c}), exhibit a more persistent stability
against non-homogeneous distribution and near-wall concentration with respect to
the larger $8~\mu m$ particles. The larger particles tend to form clusters in the core region
and accumulate at the wall at higher rates acting as an additional thermal resistance both between
the walls and the fluid and between the fluid and the particles: as a result of this behavior,
the heat transfer is reduced.

\section{Conclusions and future development}
Heat transfer enhancement is a fascinating subject with extremely interesting 
possibilities for application.
One option to increase heat transfer is to devise a new concept of heat transfer media 
constituted by a base fluid in which suitably-chosen heat transfer agents, precisely 
micro and nano particles, are injected. In this way, the fluid can be a standard 
fluid characterized by simplicity of use and well-known properties, like water, and 
the heat transfer agents can be heavy-metal, high-heat-capacity,
dispersed particles (e.g. copper, gold or platinum). 

Current literature trends show the potentials of such heat transfer media, called nanofluids;
yet the complicacy and the cost of experimental methods make it hard to understand the
intricacy of the mechanisms which govern the dynamics of the turbulent heat transfer
among the fluid and the particles. Current open questions range
from the optimal size of the particles, to the optimal concentration and 
practical solutions in real applications. Further complicating effects are represented by 
the particle inertia and the particle thermal inertia,
which are additional parameters.

The present study represents a first effort made in the frame a broader project
on the numerical simulation of heat transfer in nanofluids. Our strategic object
is to investigate the heat transfer mechanisms in nanofluids
and to devise a suitable numerical methodology to analyse their behavior.
In this paper, the DNS of fluid and thermal fields is assessed against literature data (Kasagi and Iida
[\ref{Kasagi}] and Na et al. [\ref{na_han}]) for one value of the shear Reynolds number, $Re^*=150$,
and two values of the Prandtl number, $Pr=0.71$ and $Pr=3$.
Further, preliminary results obtained from DNS
of two-phase solid-liquid turbulent flow in a channel with heat transfer are presented.
Hydrodynamically fully-developed, thermally-developing flow conditions have
been considered to investigate on heat transfer modulation produced by the dispersion
of micrometer sized particles. 
Two different sets of particles are considered, which are characterized by dimensionless
inertia response times equal to $St=1.56$ and $St=6.24$ and by dimensionless thermal
response times equal to $St_T=0.5$ and $St_T=2$, respectively.
The instantaneous features of the velocity field, the temperature field
and particles dispersion are discussed from a qualitative mechanistic viewpoint
to analyse the role of particles as heat transfer enhancement agents.

Building on the results presented in this paper,
an extensive numerical work combined with modeling efforts will be carried out
to achieve a deeper understanding of the reason why nanofluids conduct heat so
effectively. For instance, the increased surface interaction between the fluid and the solid
particles at the nanoscale as possible explanation to the
increased heat transferability will be investigated.
Indeed, for a given volume of material there is a greater
number of particles as their size decreases; perhaps there is more opportunity for
the nanoparticles to conduct the heat.
Furthermore, understanding of how the molecules of a base fluid keep nanoparticles
suspended, since nanoparticles are still dramatically larger than individual
molecules needs to be investigated. In this context, the effect of Brownian forces
on the kinematics of the nanoparticles should be investigated. It would
be also of interest to investigate the magnitude of the van der Waals forces between the
particles and their effect on the nanofluids dynamics. These forces are usually
small, but they become strong (and attractive) when the distance between particles
becomes of the order of tenths of nanometers.

~\\
{\bf Acknowledgments}
Support from PRIN (under Grant 2006098584$\_$004) and from HPC
Europa Transnational Access Program (under Grants 466 and 708)
are gratefully acknowledged.

\vspace{-0.3cm}
\section*{References}
\newcounter{bean}
\begin{list}%
{\arabic{bean}.}
{\usecounter{bean}\setlength{\rightmargin}{\leftmargin}}

\item Wang, X.-Q., Mujumdar, A.S.: Heat transfer characteristics of 
 nanofluids: a review. Int. J. Therm. Sci. {\bf 46}, 1-19 (2007).
\label{ref1}

\item Daungthongsuk, W., Wongwises, S.: A critical review of convective
 heat transfer of nanofluids. Renew. Sust. Energ. Rev. {\bf 11}, 797-817 (2007).
\label{ref2}

\item Das, S.K., Choi, S.U.S., Patel, H.E.: Heat transfer in nanofluids -
 A review. Heat Transfer Eng. {\bf 27}, 3-19 (2006).
\label{ref3}

\item Xuan, Y., Li, Q.: Heat transfer enhancement of nanofuids.
 Int. J. Heat Fluid Fl. {\bf 21}, 58-64 (2000).
\label{Xuan}

\item Maiga, S.E.B, Palm, S.J., Nguyen, C.T., Roy, G., Galanis, N.: Heat
  transfer enhancement by using nanofluids in forced convection flows.
  Int. J. Heat Fluid Fl. {\bf 26}, 530-546 (2005).
\label{Maiga_palm}

\item Maxwell, J.C.: A treatise on electricity and magnetism,
  2nd ed., vol. 1, Oxford: Clarendon Press, 1881, p. 435. 
\label{maxwell}

\item Hamilton, R.L., Crosser, O.K.: Thermal conductivity of heterogeneous two component
  systems.
  Ind. Eng. Chem. Fund. , vol. 1, {\bf 3}, 187-191 (1962).
\label{Hamilton}

\item Wasp, E.J., Kenny, J.P., Gandhi, R.L.: Solid-liquid flow slurry pipeline 
  transportation.
  Series on book materials handling. Clausthal: Trans. Tech. Publications, 1977.
\label{Wasp}

\item Trisaksri, V., Wongwises, S.: Critical review of
 heat transfer characteristics of nanofluids. Renew. Sust.
 Energ. Rev. {\bf 11}, 512-523 (2007).
\label{Wongwises}

\item
 Marchioli, C., Soldati, A.:
 Mechanisms for particle transfer and segregation in turbulent boundary layer.
 J. Fluid Mech. {\bf 468}, 283-315 (2002).
\label{ms02}

\item
 Pan, Y., Banerjee, S.:
 Numerical simulation of particle interaction with wall turbulence.
 Phys. Fluids {\bf 8}, 2733-2755 (1996).
\label{pb96}

\item
 Eaton, J.K., Fessler, J.R.:
 Preferential concentration of particles by turbulence.
 Int. J. Multiphase Flow {\bf 20}, 169-209 (1994).
\label{ef94}


\item
 Soldati, A.:
 Particles turbulence interactions in boundary layers.
 Z. Angew. Math. Mech. {\bf 85}, 683-699 (2005).
\label{s05}

\item
 Bec, J., Biferale, L., Boffetta, G., Celani, A., Cencini, M., 
 Lanotte, A., Musacchio, S., Toschi, F.:
 Acceleration statistics of heavy particles in turbulence.
 J. Fluid Mech. {\bf 550}, 349-358 (2005).
\label{bee}

\item Chagras, V., Oesterl\`{e}, B., Boulet, P.: On heat transfer in gas-solid pipe flows:
  Effects of collision induced alteration of the flow dynamics.
  Int. J.  Heat Mass Tran. {\bf 48}, 1649-1661 (2005).
\label{Chagras}

\item Lakehal, D., Fulgosi, M., Yadigaroglu, G.: Direct numerical simulation
  of turbulent heat transfer across a mobile, sheared gas-liquid interface.
  J. Heat Transf. {\bf 125}, 1129-1139 (2003).
\label{Lakehal_ful}

\item Hetsroni, G., Mosyak, A., Pogrebnyak, E.: Effect of coarse particles in
 a particle laden turbulent boundary layer.
  Int. J. Multiphase Flow {\bf 28}, 1873-1894 (2002).
\label{hetsroni}

\item
 Tiselj, I., Bergant, R., Mavko, B., Bajsi{\'{c}}, Hetsroni, G.:
 DNS of turbulent heat transfer
 in channel flow with heat conduction in the solid wall.
 J. Heat Transf. {\bf 123}, 849-857 (2001).
\label{tiselj}

\item Na, Y., Papavassiliou, D.V., Hanratty, T.J.: Use of direct numerical
  simulation to study the effect of Prandtl number on temperature fields.
  Int. J. Heat Fluid Flow {\bf 20}, 187-195 (1999).
\label{na_han}

\item Kawamura, H., Abe, H., Matsuo, Y.: DNS of turbulent heat transfer
 in channel flow with respect to Reynolds and Prandtl number effects.
 Int. J. Heat Fluid Fl. {\bf 20}, 196-207 (1999).
\label{kawamura}

\item Lyons, S.L., Hanratty, T.J., McLaughlin, J.B.: Direct numerical
 simulation of passive heat transfer in a turbulent channel flow.
 Int. J. Heat Mass Tran. {\bf 34}, 1149-1161 (1991).
\label{lyons}



\item Armenio, V., Fiorotto, V.: The importance of the forces acting
 on particles in turbulent flows. Phys. Fluids {\bf 8}, 2437 (2001).
\label{armenio}

\item Rizk, M.A., Elghobashi, S.E.: The motion of a spherical particle suspended
 in a turbulent flow near a plane wall. Phys. Fluids {\bf 28}, 806--817 (1985).
\label{re}

\item Soltani, M., Ahmadi, G.: Direct numerical simulation of particle entrainment
  in turbulent channel flow.
  Phys. Fluids A {\bf 7}, 647-657 (1995).
\label{Soltani}

\item Crowe, C., Sommerfeld, M., Tsuji, Y.: Multiphase Flows with Droplets
 and Particles. New York: CRC Press 1998, p. 469.
\label{crowebook}

\item Schiller, V.L., Naumann, A.: Uber die grundlegenden Berechnungen
  bei der Schwerkraftaufbereitung.
  Z. Ver. Deut. Ing. {\bf 77}, 318-320 (1935).
\label{Schiller}

\item W. Ranz and W. Marshall:
 Evaporation from Drops. Chem. Eng. Progress,
 {\bf 48}, 142-180 (1952).
\label{rm}



\item Boivin, M., Simonin, O., Squires, K.D.: Direct numerical simulation
 of turbulence modulation by particles in isotropic turbulence.
 J. Fluid Mech. {\bf 375}, 235-263 (1998).
\label{Boivin}

\item Sundaram, S., Collins, L.R.: A numerical study of the modulation
 of isotropic turbulence by suspended particles.
  J. Fluid Mech. {\bf 379}, 105-143 (1999).
\label{Sundaram}

\item Lam, K., Banerjee, S.: On the condition of streak formation in bounded
  flows.
  Phys. Fluids A {\bf 4}, 306-320 (1992).
\label{Lam1}

\item Kasagi, N. and Iida, O.: Progress in direct numerical simulation of 
 turbulent heat transfer.
 Proceedings of the 5th ASME/JSME Joint Thermal Engineering Conference,
 In CD-ROM, March 15-19, San Diego, California (1999).
\label{Kasagi}

\item Uijttewaal, W.S., Oliemans, R.V.A: Particle dispersion and deposition in direct numerical and
  and large eddy simulations of vertical pipe flow.
  Phys. Fluids {\bf 8}, 2590-2604 (1996).
 \label{Uijttewaal}

\item Kiger, K.T., Pan, C.: Suspension and turbulence modification effects
  of solid particulates on a horizontal turbulent channel flow.
  J. Turbul. {\bf 3}, 1-21 (2002).
\label{Kiger}


\item Monin, A.S., Yaglom, A.M.: Statistical Fluid Mechanics:Mechanism of 
  Turbulence,
  book 2, MIT Press, 1975, p. 873.
\label{Monin}

\item Batchelor, G.K.: Small-scale variation of convective quantities like temperature
  in turbulent fluid.
  J. Fluid Mech. A {\bf 5}, 113-133 (1959).
\label{Batchelor}

\item Shotorban, B., Mashayek, F., Pandya, R.V.R.: Temperature statistics in
  particle-laden turbulent homogeneous shear flow.
  Int. J. Multiphase Flow {\bf 29}, 1333-1353 (2003).
\label{shotorban}

\item Jaberi, F.A., Mashayek, F.: Temperature decay
  in two-phase turbulent flows.
  Int. J. Heat Mass Tran. {\bf 43}, 993-1005 (2000).
\label{jaberi}


%
\label{Roy}

\item van Haarlem, B., Boersma, B.J., Nieuwstadt, F.T.M : Direct numerical simulation
 of particle deposition onto a free-slip and no-slip surface. 
 Phys. Fluids A. {\bf 10}, 2608-2620 (1998).
\label{haarlem}

\item Portela, L.M., Cota, P., Oliemans, R.V.A : Numerical study of the near wall
 behaviour of particles in turbulent pipe flows.
 Powder Technol. {\bf 125}, 149-157 (2002).
\label{portela}

\item Reeks, M.W.: The transport of discrete particles in 
 inhomogeneous turbulence.
 J. Aerosol Sci. {\bf 310}, 729-739 (1983).
\label{reeks}

\item Adrian, R.J., Meinhart, C.D., Tomkins, C.D.:
Vortex organization in the outer region of
the turbulent boundary layer.
J. Fluid Mech. {\bf 422}, 1-54 (2000).
\label{adrian}

\end{list}





\end{document}